\documentclass[10pt, oneside]{article}
\pdfoutput=1
\usepackage{hyperref}
\usepackage{enumitem} 
\usepackage{easylist}  	
\usepackage{graphicx}			
\usepackage{amssymb}
\usepackage[usenames]{color}
\usepackage{epsf}
\usepackage[usenames,dvipsnames]{pstricks}
\usepackage{epsfig}
\usepackage{pst-grad} 
\usepackage{url}
\usepackage{wrapfig}
\usepackage{float} 
\usepackage[abs]{overpic} 
\newenvironment{s-enumerate}{
\begin{enumerate}
  \setlength{\itemsep}{1pt}
  \setlength{\parskip}{0pt}
  \setlength{\parsep}{0pt}
}{\end{enumerate}}

\newenvironment{s-itemize}{
\begin{itemize}
  \setlength{\itemsep}{1pt}
  \setlength{\parskip}{0pt}
  \setlength{\parsep}{0pt}
}{\end{itemize}}

\let\OLDthebibliography\thebibliography
\renewcommand\thebibliography[1]{
  \OLDthebibliography{#1}
  \setlength{\parskip}{.5ex}
  \setlength{\itemsep}{0pt plus 0.3ex}
}
\title{Minimal Glider-Gun in a 2D Cellular Automaton}
\author{Jos\'e Manuel G\'omez Soto%
\thanks{jmgomezuam@gmail.com, http://matematicas.reduaz.mx/$\sim$jmgomez}%
\hspace{2ex}{\it \small Universidad Aut\'onoma de Zacatecas.}\\
\hspace{2ex}{\it \small  Unidad Acad\'emica de Matem\'aticas. Zacatecas, Zac. M\'exico.}\\
Andrew Wuensche%
\thanks{andy@ddlab.org,  http://www.ddlab.org}%
\hspace{2ex}{\it \small Discrete Dynamics Lab.}\\
\\
}

\begin{document}

\maketitle

\vspace{-3ex}
\begin{abstract}

\noindent To understand the underlying principles of self-organisation
and computation in cellular automata, it would be helpful to find the
simplest form of the essential ingredients, glider-guns and eaters,
because then the dynamics would be easier to interpret. Such minimal components
emerge spontaneously in the newly discovered Sayab-rule, a binary 2D
cellular automaton with a Moore neighborhood and isotropic
dynamics. The Sayab-rule has the smallest glider-gun
reported to date, consisting of just four live cells at its minimal
phases.  We show that the Sayab-rule can implement complex dynamical
interactions and the gates required for logical universality.

\end{abstract}

\begin{center}
{\it keywords: universality, cellular automata, glider-gun, logical gates.}
\end{center}

\section{Introduction}
\label{Introduction}

\noindent The study of 2D cellular automata (CA) with complex
properties has progressed over time in a kind of regression from the
complicated to the simple. Just to mention a few key moments in
CA history, the original CA was von~Neumann's with 29 states designed to
model self-reproduction, and by extension -- universality\cite{von Neumann}.
Codd simplified von~Neumann's CA to 8 states\cite{Codd68}, and Banks
simplified it further to 3 and 4 states\cite{Banks70,Banks71}. In
modelling self-reproduction its also worth mentioning Langton's
``Loops''\cite{Langton84} with 8 states, which was simplified 
by Byl to 6 states\cite{Byl89}.
These 2D CA all featured the 5-cell ``von~Neumann'' neighborhood.

Another line of research was based on the larger 9$\times$9 ``Moore''
neighborhood.  Conway's famous ``Game-of-Life'' binary CA\cite{Berlekamp1982,Gardner1970}
featured the first emerging gliders, and Gosper was able to devise ``glider-guns''
to fire a stream of gliders. Interactions involving glider-streams and
``eaters'' enabled the demonstration of universal computation.  A few
``Life-Like'' CA featuring glider-guns were subsequently discovered
that follow the Game-of-Life birth/survival
paradigm\cite{Eppstein2010}.

\begin{figure}[htb]
\begin{center}
\begin{minipage}[c]{.55\linewidth} 
\begin{minipage}[c]{.2\linewidth}
  \includegraphics[width=1\linewidth, bb=32 102 58 132, clip=]{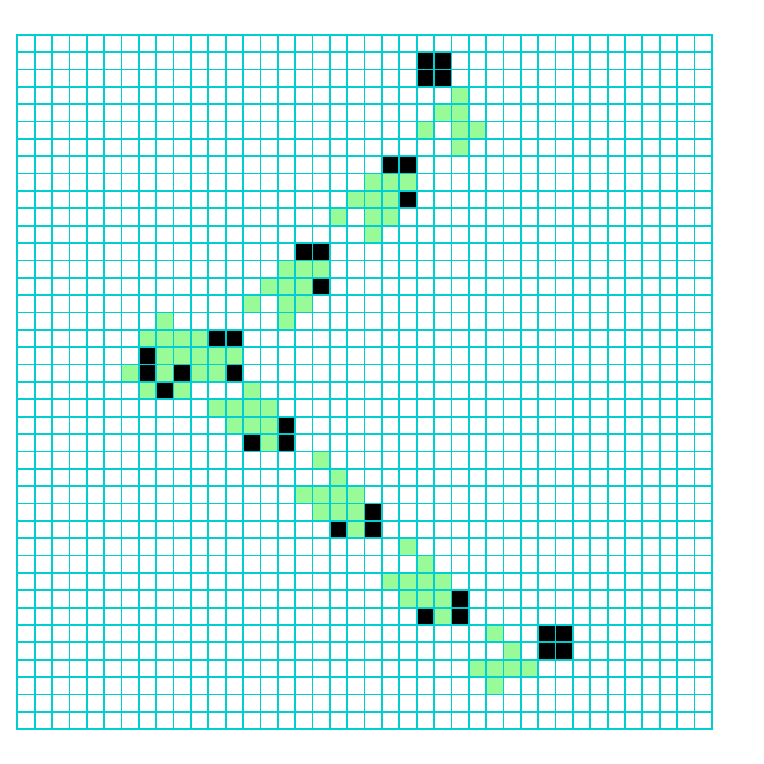}
\end{minipage}
\hfill
\begin{minipage}[c]{.5\linewidth}
  \includegraphics[width=1\linewidth, bb=26 12 169 205, clip=]{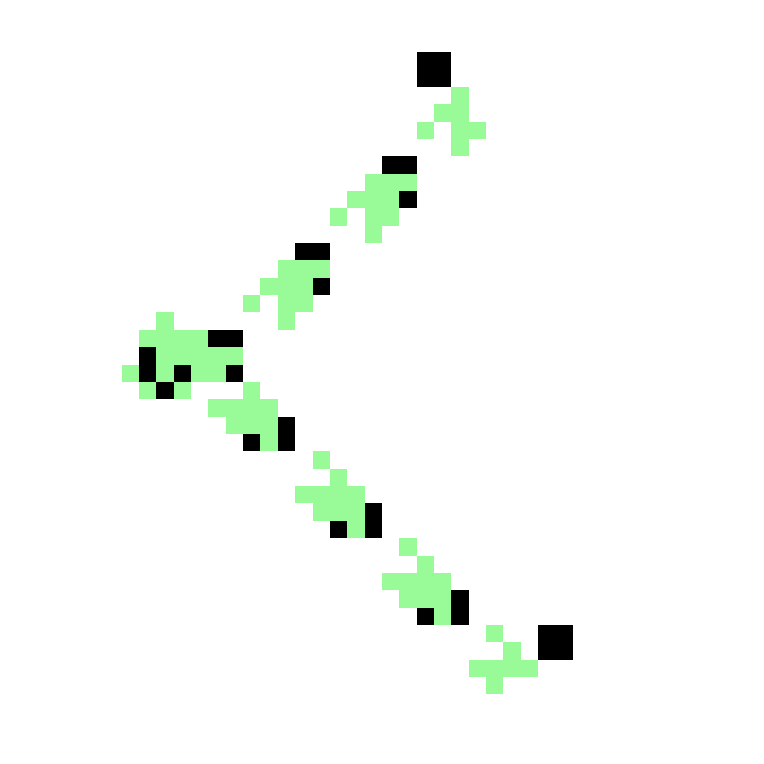}
\end{minipage}
\end{minipage}
\end{center}
\vspace{-4ex}
\caption[basic glider-gun GG1]
{\textsf{\underline{\it Left}: One of the Sayab-rule's minimal glider-gun patterns, 
of 4 live cells. \underline{\it Right}: 
the glider-gun GG1 in action shooting two diagonal glider streams 
with a frequency of 20 time-steps and glider spacing of 5 cells.
Each glider streams is stopped by an eater. Because the system is isotropic, any orientation
of the glider-gun is equally valid. Green dynamic trail are set to 10 time-steps.\\
\small{Note: Green dynamic trails mark any change
on a zero (white) cell within the last 10 time-steps, giving a glider a green trailing wake.
10 time-steps is the setting in all subsequent figures with green dynamic trails.}}}
\label{glider-gun GG1}
\end{figure}

\begin{figure}
\begin{minipage}[c]{.22\linewidth}
\textcolor{white}{x\\ x\\ x\\ x\\ x\\ x\\}
\includegraphics[width=1\textwidth, bb= 2 2 166 122]{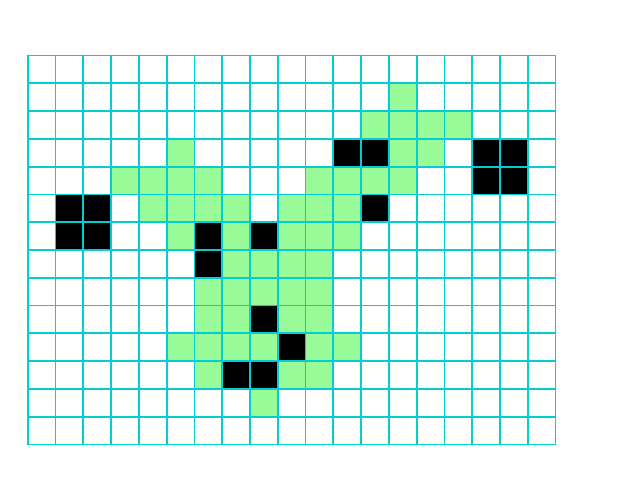} 
\end{minipage}
\hfill
\begin{minipage}[c]{.79\linewidth}
\hspace{-1ex}\includegraphics[width=1\textwidth]{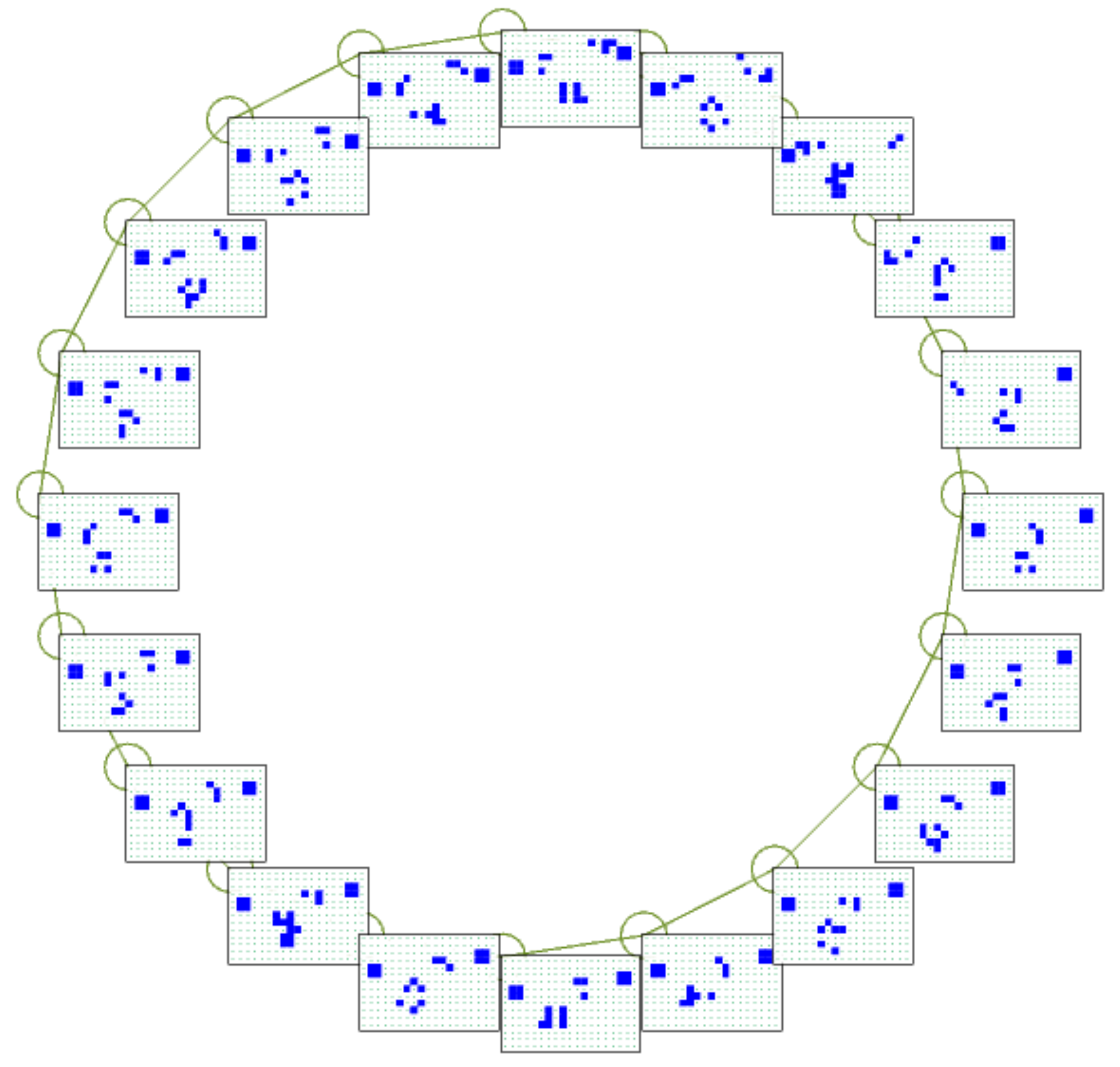}
\end{minipage}
\vspace{-4ex}
\caption{
{\textsf{The Sayab-rule glider-gun attractor cycle\cite{Wuensche92} 
with a period of 20 time-steps composed of two phases, where 
opposite glider-gun patterns are flipped. The direction of time is clockwize.
A small patch was isolated around a glider-gun by two close eaters.
\underline{\it Left}: A detail of a patch with a minimal glider-gun
(green denotes change) alongside the same pattern on the attractor cycle.}}}
\label{cycle-state}
\end{figure}
\clearpage

More recently, CA that feature glider-guns, but not based on
birth/survival, have been found, including Sapin's
R-Rule\cite{Sapin2004}, and the authors' X-Rule\cite{Gomez2015} and
Precursor-Rule\cite{Gomez2017}.  Glider-guns have also been discovered
in CA with 6 and 7 cell neighborhoods on a hexagonal 2D geometry with
3 values\cite{Wuensche05,Wuensche2006}. From this we can see that the
architecture of CA that is demonstrably able to support emerging
complex dynamics is becoming simpler --- arguably a positive
development since a minimal system becomes easier to interpret. This is
important if the underlying principles of universal computation in
CA are to be understood, and by extension the
underlying principles of self-organisation in nature.

The essential ingredients for a recipe to create logical universality
in CA are gliders, glider-guns, eaters, and the appropriate diversity
of dynamical interactions between them including bouncing and
destruction. Of these the glider-gun or ``pulse generator'', a devise
that ejects gliders periodically, is the most critical and elusive
structure.  To some extent glider-guns have been demonstrated in
1D\cite{Cook2004}, and to an lesser extent in 3D\cite{Wuensche-3D-GG},
but here we consider the more familiar and much more studied 2D space,
which is also easier to represent and manipulate.  Up to now,
glider-guns in 2D CA comprise periodic structures that involve at
least tens of cells in the $on$ state in their minimum phase.
Here we present a much smaller glider-gun which emerges spontaneously in the
newly discovered Sayab-rule, named after the Mayan-Yucatec word for a
spring (of running water).

The Sayab-rule is a binary 2D CA with a Moore neighborhood and
isotropic dynamics. Though analogous to the game-of-Life and the
recently discovered Precursor-rule, the Sayab-rule has the smallest
glider-gun reported to date, consisting of just four live cells at its
minimal phase, as well as eaters and other essential ingredients.  We
show that the Sayab-rule can implement a diversity of complex
dynamical structures and the logical gates required for logical
universality\footnote{We designate a CA ``logically universal'' if its
  possible build the logical gates NOT, AND, and OR, to satisfy
  negation, conjunction and disjunction.  ``Universal computation'' as
  in the Game-of-Life requires additional
  functions\cite{Randall2002,Berlekamp1982}, memory registers, auxiliary storage
  and other components.}, and supports analogous complex structures
from the Game-of-Life lexicon --- still lives, eaters, oscillators and
spaceships.

The paper is organised into the following further sections,
(\ref{The Sayab-Rule definition}) the Sayab-Rule definition,  
(\ref{Glider-guns, eaters and collisions}) the Sayab-Rule's gliders-guns,
eaters, collisions, and other complex structures,
(\ref{Logical Universality and Logical Gates}) logical universality by logical gates, and
(\ref{Concluding remarks})  the concluding remarks.

\section{The Sayab-Rule definition}
\label{The Sayab-Rule definition}

\begin{figure}[htb]
\begin{center}
  \includegraphics[width=.8\linewidth, bb=2 9 544 77]{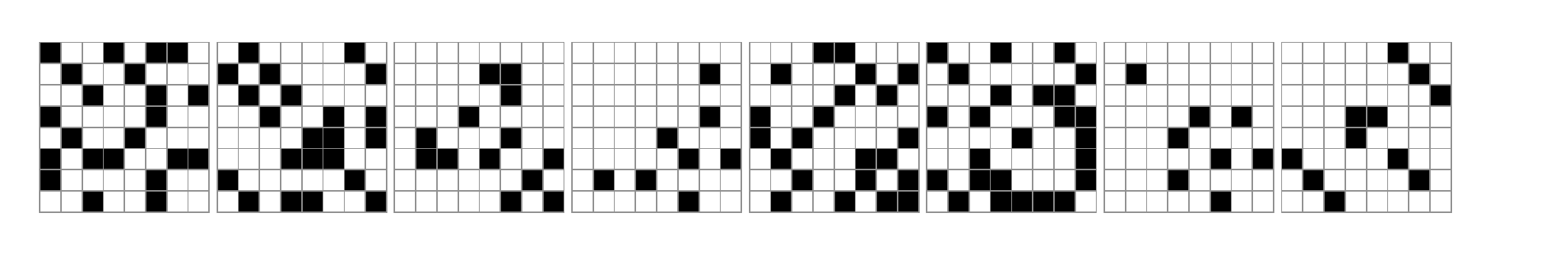}\\[2ex]
\includegraphics[width=1\linewidth]{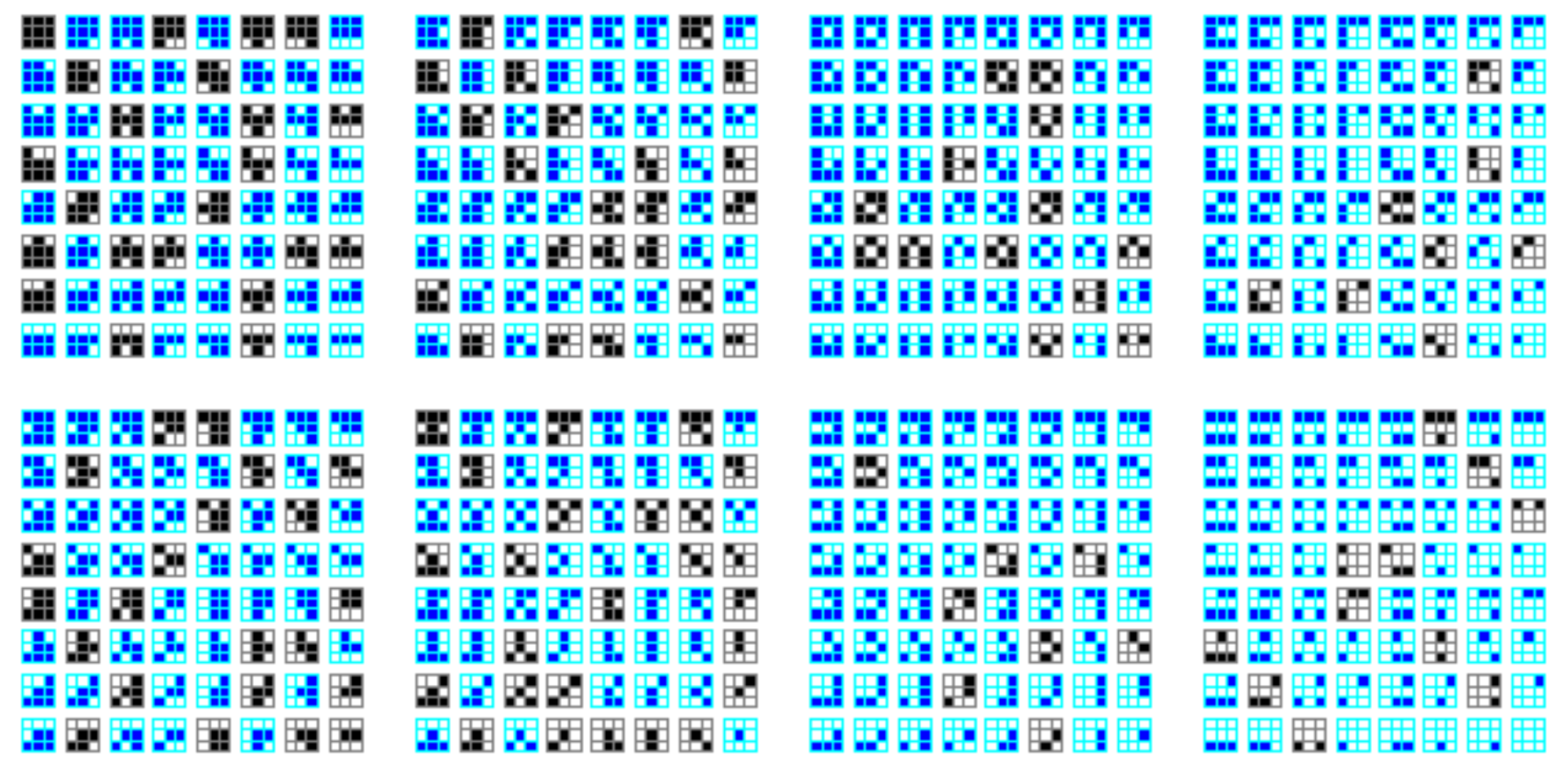}
\end{center}
\vspace{-4ex}
\caption[Sayab Rule]
{\textsf{\underline{\it Top} The Sayab rule-table based on to all 512 neighborhoods, 
and \underline{\it Below} expanded to show each neighborhood pattern.
131 black neighborhoods map to~1, 381 blue neighborhoods map to 0. Because the rule
is isotropic, only 102 symmetry classes are significant, as described in 
figure~\ref{isotropic neighborhoods with output=1}}}\label{lockup table}
\label{sayab-ruletable}
\end{figure}

\begin{figure}[htb]
\begin{center}
\includegraphics[width=.95\textwidth]{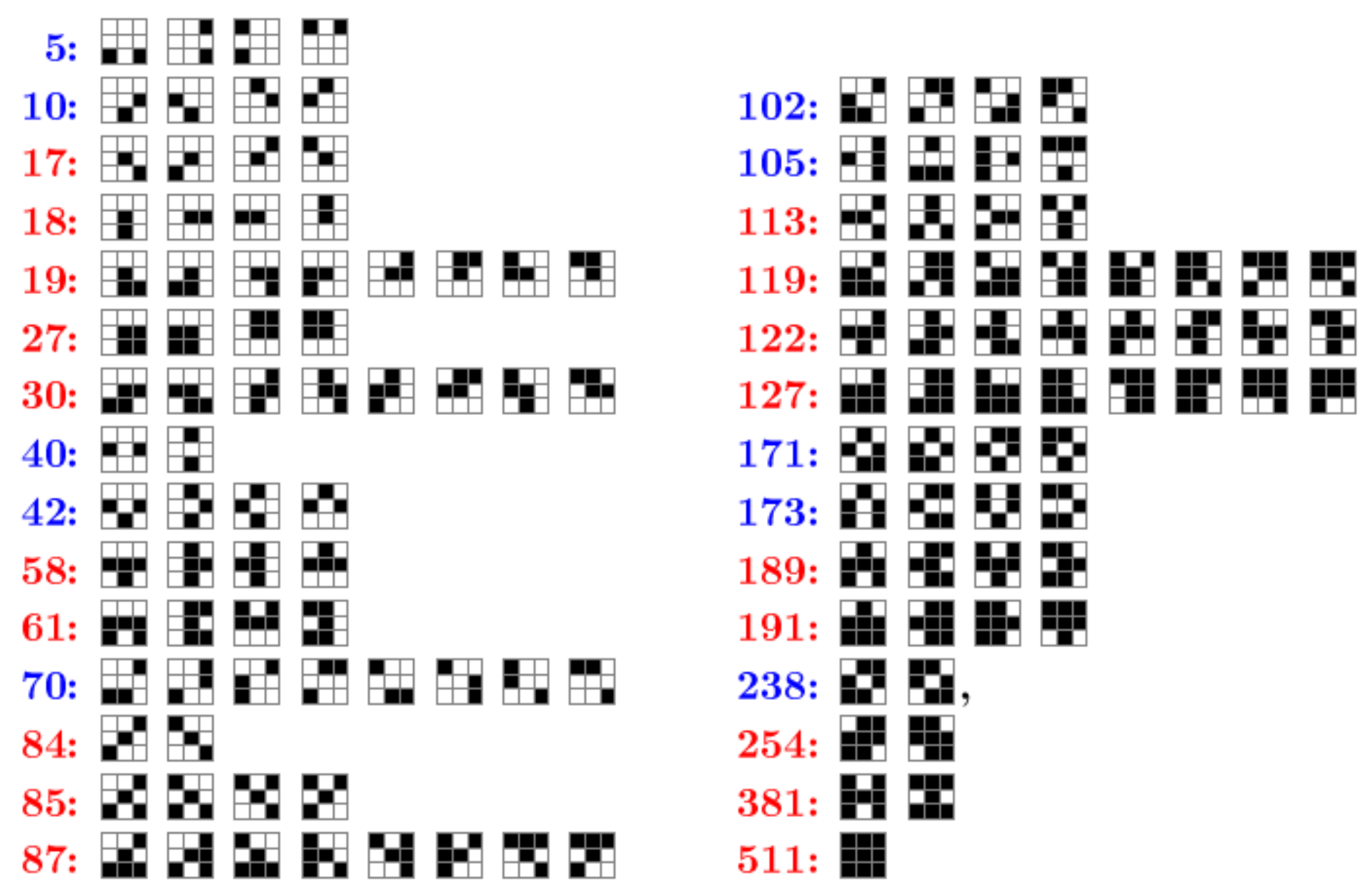}
\end{center}
\vspace{-4ex}
\caption[isotropic neighborhoods with output=1]%
{\textsf{The Sayab-rule's 29 isotropic neighborhood symmetry classes that map to~1
(the remaining 73 symmetry classes map to~0, making 102 in total).
Each class is identified by the smallest decimal equivalent of the class, where the 3$\times3$
pattern is taken as a string in the order 
\begin{minipage}[c]{4ex}\scriptsize 876\\[-1ex]543\\[-1ex]210\end{minipage} --- for example, 
  the pattern
  \raisebox{-1.4ex}{\includegraphics[width=2.8ex,bb=7 4 23 25, clip=]{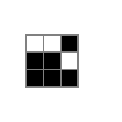}} 
is the string 001110111 representing the symmetry class 119. The class numbers are
colored depending on the value of the central cell to
distinguish birth (blue) from survival (red), but no clear ``Life-like''
birth/survival logic is discernible.  
}}
\label{isotropic neighborhoods with output=1}
\end{figure}

The Sayab-Rule is found in the ordered region of the input-entropy
scatter-plot\cite{Wuensche99} close to the Precursor
Rule\cite{Gomez2017}, and from the same sample and
short-list\cite{Gomez2015,Gomez2017}.  The input-entropy criteria in
this sample followed ``Life-Like'' constraints (but not birth/survival
logic) to the extent that the rules are binary, isotropic, with a
Moore neighborhood, and with the $\lambda$~parameter\cite{Langton90}, the density of
1s in the look-up table, similar to the Game-of-Life where
$\lambda=0.273$.  Isotropic mapping --- the same output for any
neighborhood rotation, reflection or vertical flip --- reduces the
full rule-table (figure~\ref{sayab-ruletable}) with $2^9=512$
neighborhood outputs to just $102$ effective outputs\cite{Sapin2010},
from which just 29 ``symmetry classes'' map to~1 
(figure~\ref{isotropic neighborhoods with output=1}).

\section{Glider-guns, eaters and collisions}
\label{Glider-guns, eaters and collisions}

From the game-of-Life lexicon, we borrow the various names for
characteristic patterns or objects, including glider-guns, gliders,
eaters, still-lives, oscillators, and space-ships.  A glider is a
periodic mobile pattern that recovers its shape but at a displaced
position, making it move at a given velocity, sometimes referred to as
a mobile particle. A glider is usually identified as moving on the
diagonal, whereas an orthogonal ``glider'' is called a space-ship.  A
glider-gun is a periodic pattern in a fixed location that sends,
shoots, or sheds, gliders into space at regular intervals.
 
In the Sayab-rule, the spontaneous emergence of its basic glider-gun,
as well as isolated gliders, is highly probable from a sufficiently
large random initial state because the four glider patterns are very
simple and likely to occur or emerge by chance -- likewise, the
smallest glider-gun patterns.  Simple still-lives and oscillators
(which may act as eaters which destroy gliders but remain active) are
also likely to occur or emerge from random patterns. The basic
glider-gun is also probable in subsequent evolution because it can
result from the collision of two gliders, or a glider and an
oscillator, though the glider-gun can also be destroyed by incoming
gliders and other interactions.

\vspace{-3ex}
\begin{figure}[h]
\textsf{\small
\begin{center}
  \begin{tabular}[t]{ c c c c c c }
 \multicolumn{4}{c}{$\leftarrow$------------------ Ga -----------------$\rightarrow$} &\\
   \includegraphics[height=.08\linewidth,bb=-1 -2  42 35,  clip=]{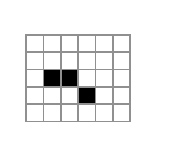}%
 & \includegraphics[height=.08\linewidth,bb=-3 -2  42 35,  clip=]{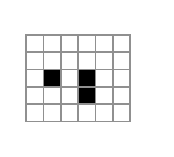}%
 & \includegraphics[height=.08\linewidth,bb=-3 -2  42 35,  clip=]{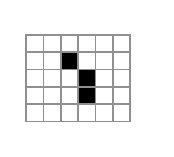}%
 & \includegraphics[height=.08\linewidth,bb=-3 -2  42 35,  clip=]{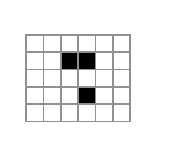}%
 & \includegraphics[height=.08\linewidth,bb=-3 -2  42 35,  clip=]{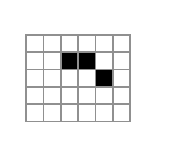}
 & \includegraphics[height=.08\linewidth,bb= 0 -2 30 30, clip=]{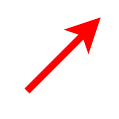}
\\[-2ex]  
1 & 2 & 3 & 4 & 5 
\end{tabular}
\end{center}
}
\vspace{-4ex}
\caption[glider Ga]%
{\textsf{The 4 phases of the Sayab-rule glider Ga, moving NE with speed $c$/4,
where $c$ is the ``speed of light'', in this case, for a Moore neighborhood,
$c$ equals one cell per time-step, diagonally or orthogonally.
}}
\label{glider-Ga}
\vspace{-3ex}
\end{figure}

\begin{figure}[h]
\begin{center} 
\begin{minipage}[c]{.48\linewidth}
  \includegraphics[width=1\linewidth, bb= 0 -7 240 62, clip=]{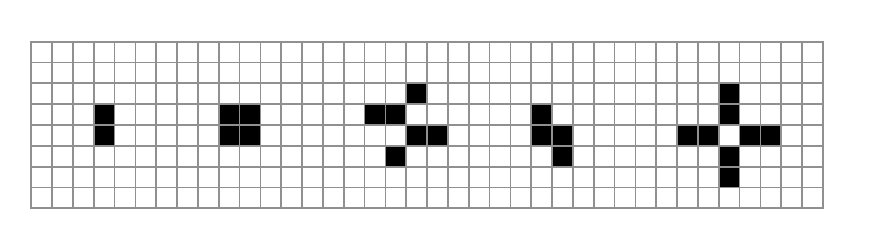}
\end{minipage}
\end{center}
\vspace{-5ex}
\caption[Examples of still-lives]
{\textsf{Examples of still-lives.}}
\label{still-lives}
\end{figure}

\begin{figure}[htb]
\begin{center} 
\begin{minipage}[c]{.8\linewidth}
  \includegraphics[width=1\linewidth,bb= 33 189 361 238, clip=]{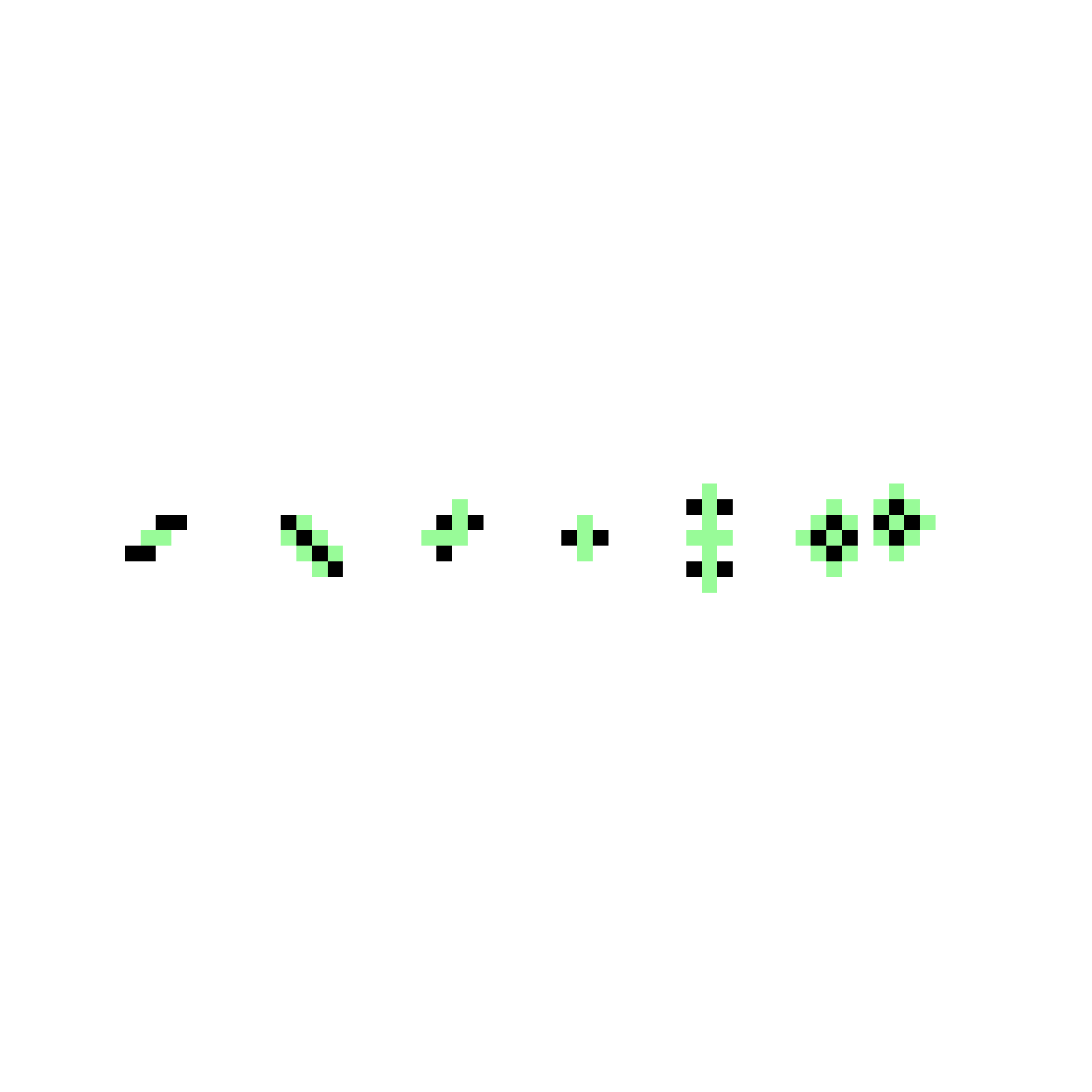}\\[-1ex]
\color{white}--\color{black} $p$=2
\color{white}----------\color{black} $p$=2
\color{white}--------\color{black} $p$=4
\color{white}------\color{black} $p$=4
\color{white}------\color{black} $p$=4
\color{white}---------\color{black} $p$=9

\end{minipage}
\end{center}
\vspace{-3ex}
\caption{Sayab rule oscillators with the periods indicated.}
\label{oscillators}
\end{figure}

\begin{figure}[htb]
\begin{center}
  \fbox{\includegraphics[width=.8\linewidth,bb=78 79 363 357, clip=]{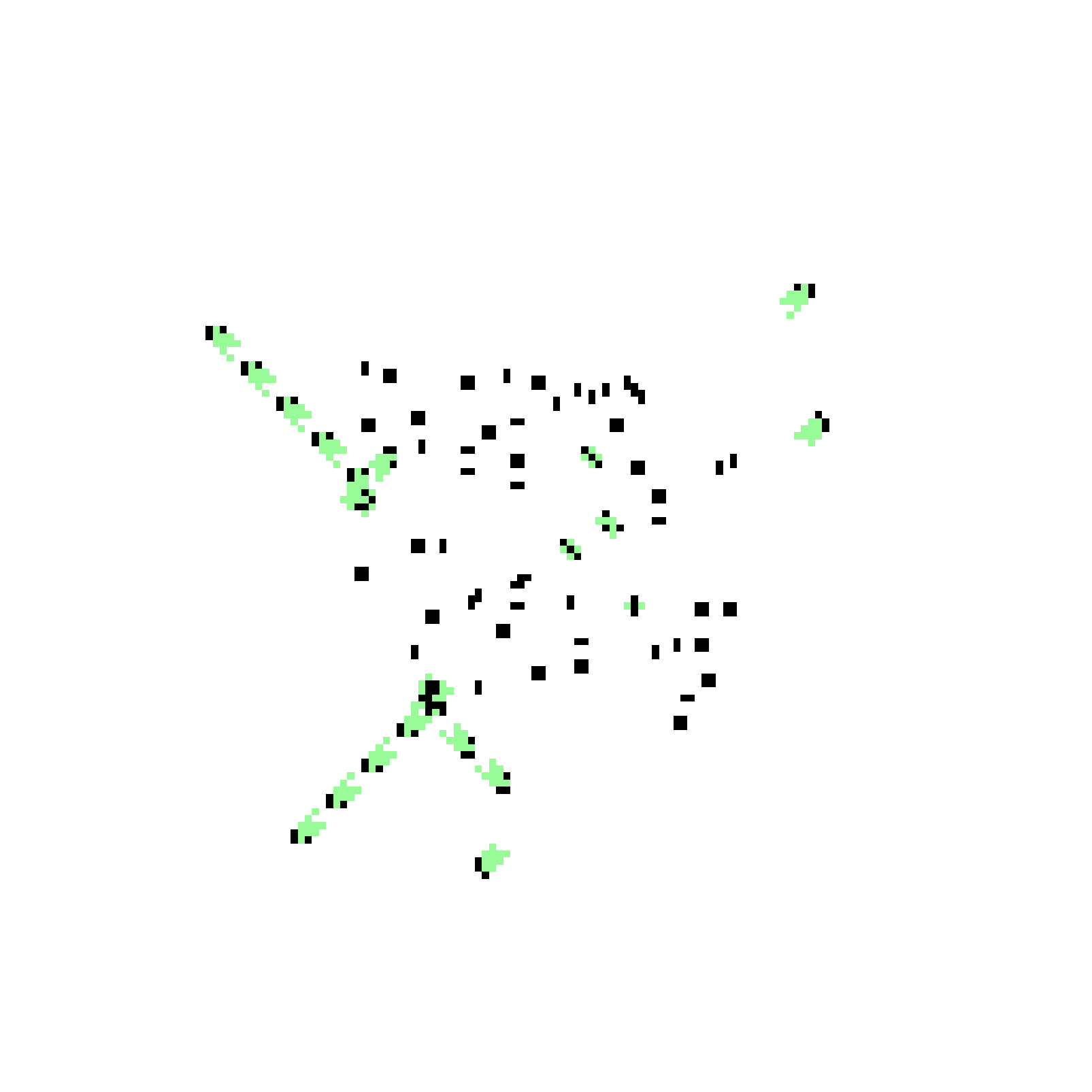}} 
\end{center} 
\vspace{-3ex}
\caption[Typical evolution from a random zone]
{\textsf{
A typical evolution emerging after 108 time-steps from a 50x50
30\% density random zone. 
Two stable glider-guns have emerged, together with other gliders, still-lives and oscilators.
\label{Typical evolution}
}}
\end{figure}

\begin{figure}[htb]
\begin{center}
  \includegraphics[width=1\linewidth,bb= 0 -2 543 78, clip=]{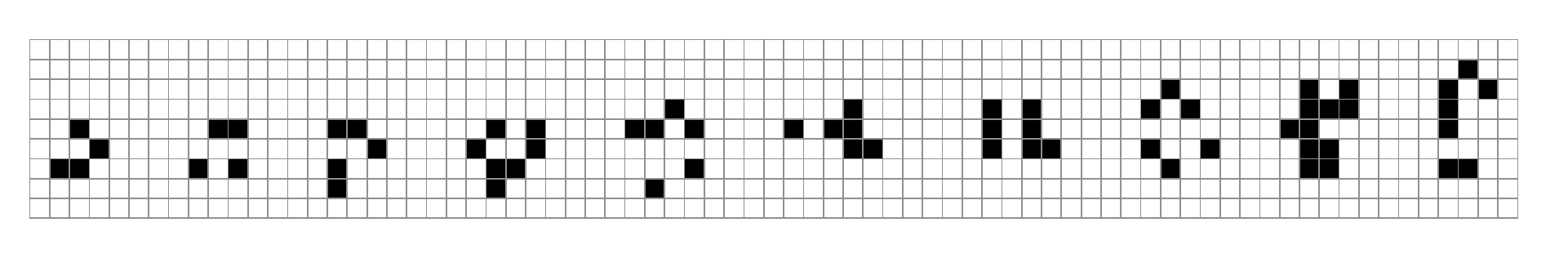} 
\end{center}
\vspace{-5ex}
\caption{
{\textsf{The glider-gun core for 10 successive time-steps --- in the next
next 10 time-steps the same reversed patterns are repeated,
to make the period 20 attractor cycle (figure~\ref{cycle-state}). The pattern
sequence is from left to right. Any of these patterns are the seeds of a glider-gun,
with the smallest, 4 live cells, being the most probable to occur in a random pattern.
}}}
\label{gg-core10}
\end{figure}

\begin{figure}[h]
\begin{center}
\begin{minipage}[c]{.7\linewidth}
  \begin{overpic}[width=.45\linewidth, bb=69 106 236 331, clip=]{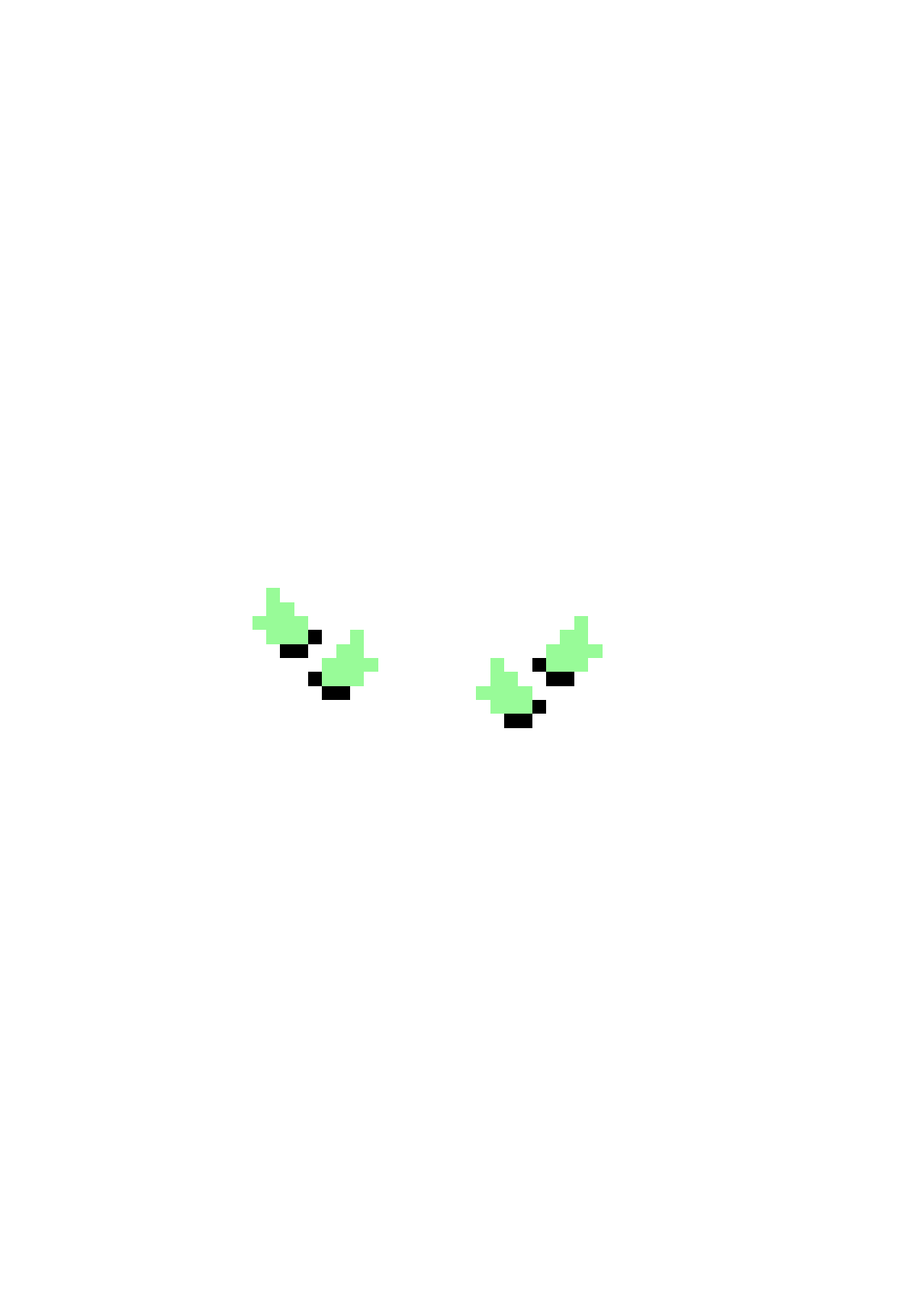} 
\put(50,5){(a)}
\end{overpic}
\hfill
\begin{overpic}[width=.45\linewidth, bb=69 106 236 331, clip=]{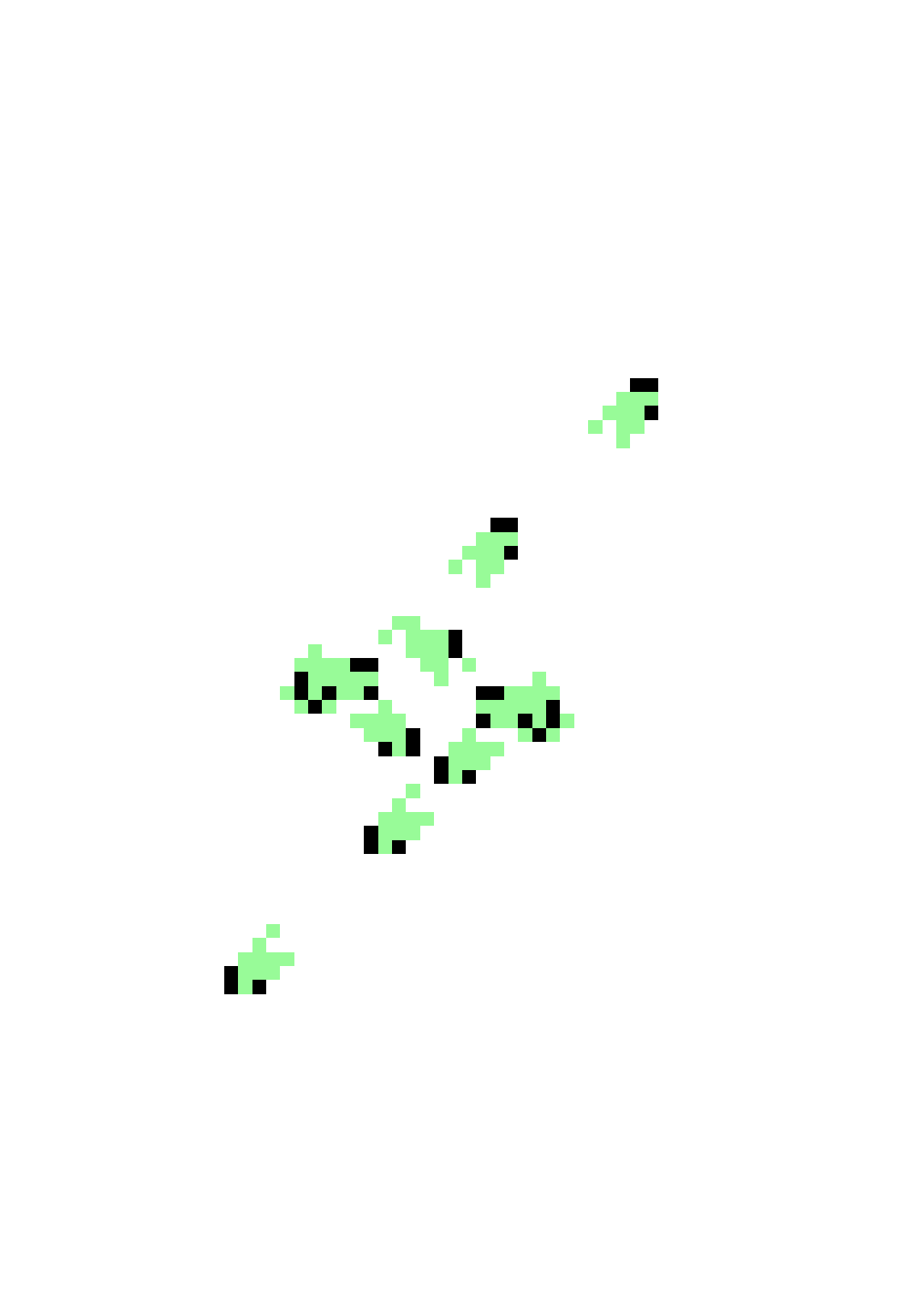}
\put(50,5){(b)}
\end{overpic}
\end{minipage}
\end{center}
\vspace{-4ex}
\caption[compound glider-gun GG2]
{\textsf{(a) two pairs of gliders, each pair colliding at 90$^{\circ}$,
form a pre-image of GG2. (b) the compound glider-gun GG2 shown after 138 time-steps,
shoots gliders with a frequency of 40 time-steps and glider spacing is 10 cells.}}
\vspace{-3ex}
\label{GG2}
\end{figure}

\begin{figure}[h]
\begin{center}
 \includegraphics[width=.6\linewidth, bb=4 39 392 336, clip=]{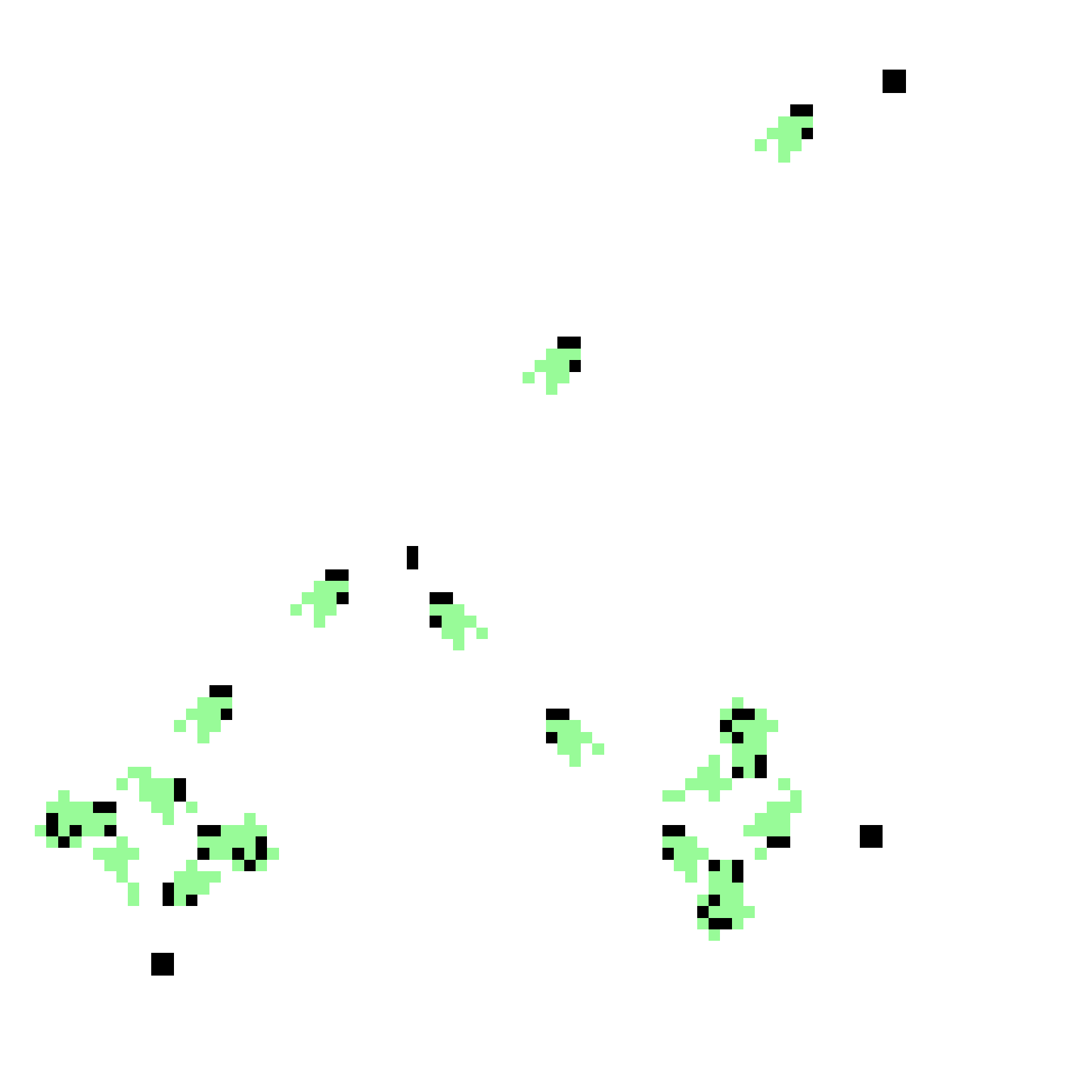} 
\end{center}
\vspace{-4ex}
\caption{
{\textsf{The compound glider-gun GG4 shoots gliders with a frequency of
80 time-steps and glider spacing is 20 cells.}}}
\label{GG4}
\end{figure}

As can be seen in its attractor\cite{Wuensche92} (figure~\ref{cycle-state}), the
Sayab-rule's basic glider-gun GG1 (figure~\ref{glider-gun GG1})
has a core that varies between just 4 and 11 live cells
during its cycle of twenty time-steps, which is composed of two
equivalent phases of 10 time-steps. After 10 time-steps the core
patters are reversed. In figures~\ref{cycle-state} and
\ref{gg-core10} the core and its twin 45$^{\circ}$ glider streams face
towards the North, but the glider-gun can be oriented to face in any
of 4 directions.  The glider-gun shoots gliders at 20 time-steps
intervals with a speed is $c$/4, and a glider takes 20 time-steps to
traverse 5 (diagonal) cells, which is also the spacing of
gliders in a glider stream. This spacing can be doubled (without limit)
by combining the basic glider-guns into compound glider-guns 
(figures~\ref{GG2} and \ref{GG4}).

In the Sayab rule, there are many possible outcomes resulting
from collisions between two (or more) gliders, and between gliders and
still-lives or oscillators. These have been examined experimentally
but not exhaustively.
The outcomes depend on the precise timing and
points of impact, and can result in the destruction, survival, or
modification of the various colliding objects.  For the purposes of
this paper we highlight some significant collision outcomes.

\begin{figure}[htb]
\begin{center}
\begin{minipage}[c]{.4\linewidth}
  \begin{overpic}[width=.27\linewidth, bb=203 70 249 105, clip=]{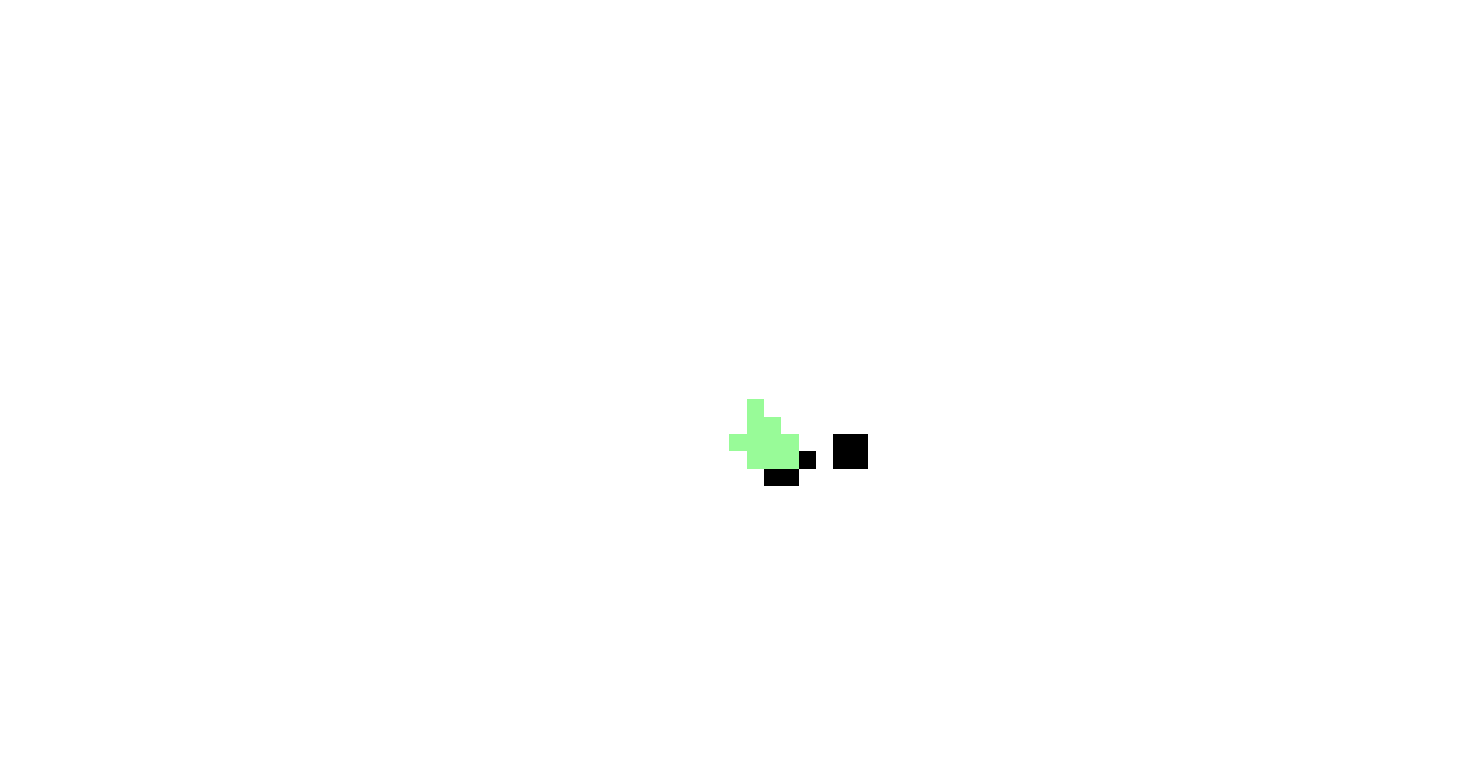}
\put(-10,5){(a)}
\end{overpic}
\hfill
\begin{overpic}[width=.3\linewidth, bb=220 60 273 100, clip=]{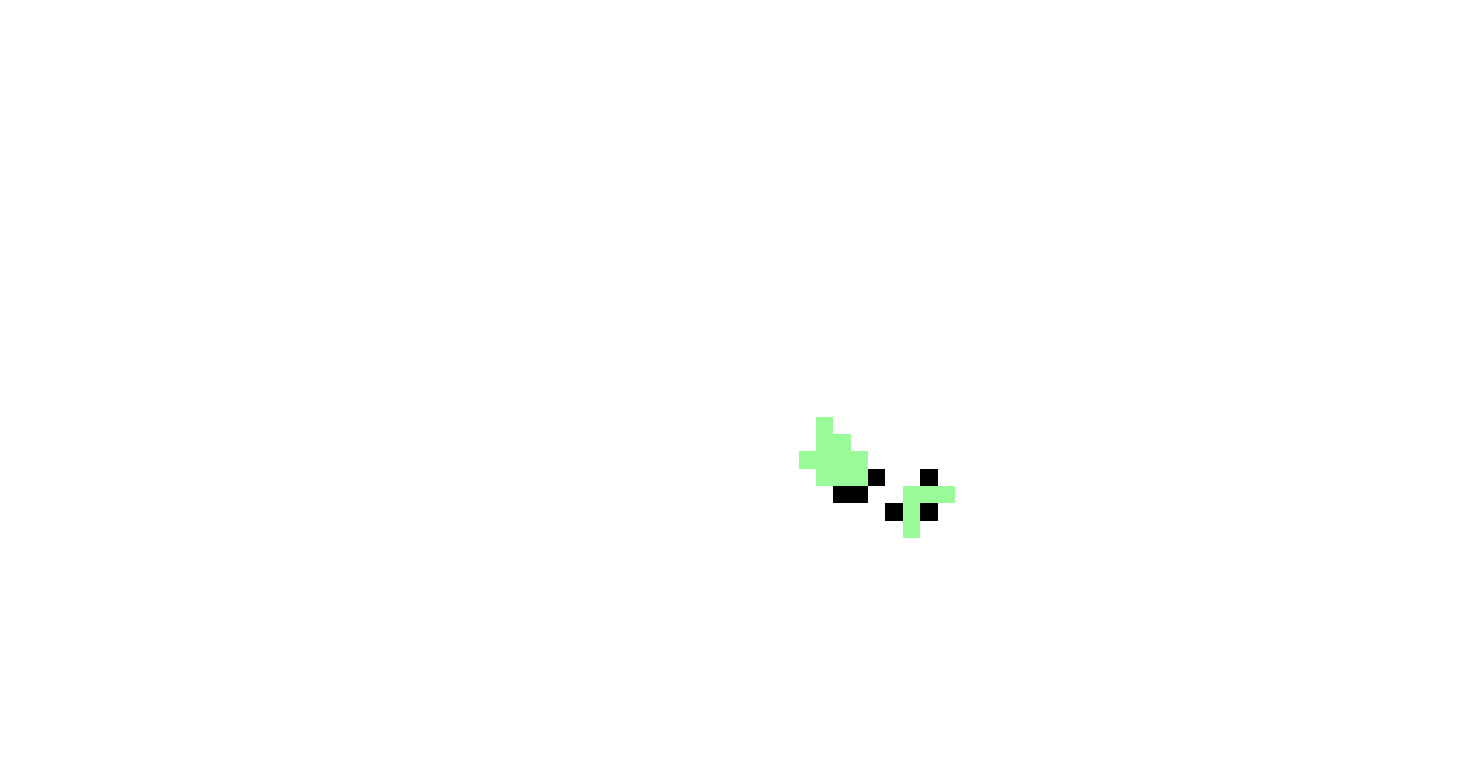}
\put(-10,5){(b)}
\end{overpic}
\end{minipage}
\end{center}
\vspace{-4ex}
\caption[Eaters]
{\textsf{Collisions between a glider and an eater, (a) derived from a still-life, and 
(b) from an oscillator.}}
\label{eater1}
\end{figure}

\begin{figure}[htb]
\begin{center}
\begin{minipage}[c]{.55\linewidth}
  \begin{overpic}[width=.3\linewidth, bb=40 51 90 217, clip=]{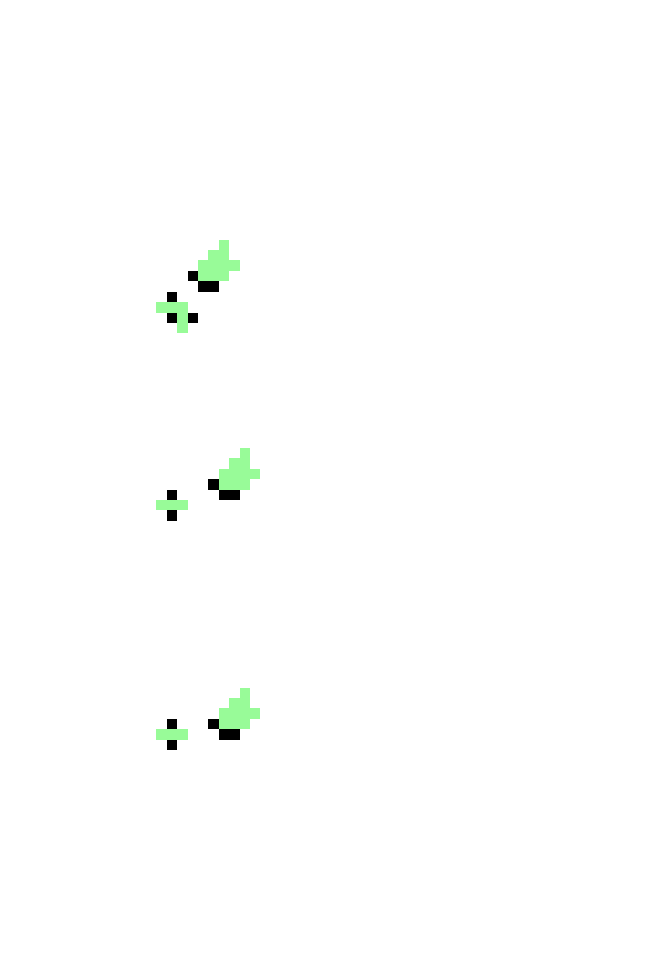}
\put(-10,5){(a)}
\end{overpic}
\hfill
\begin{overpic}[width=.3\linewidth, bb=40 51 90 217, clip=]{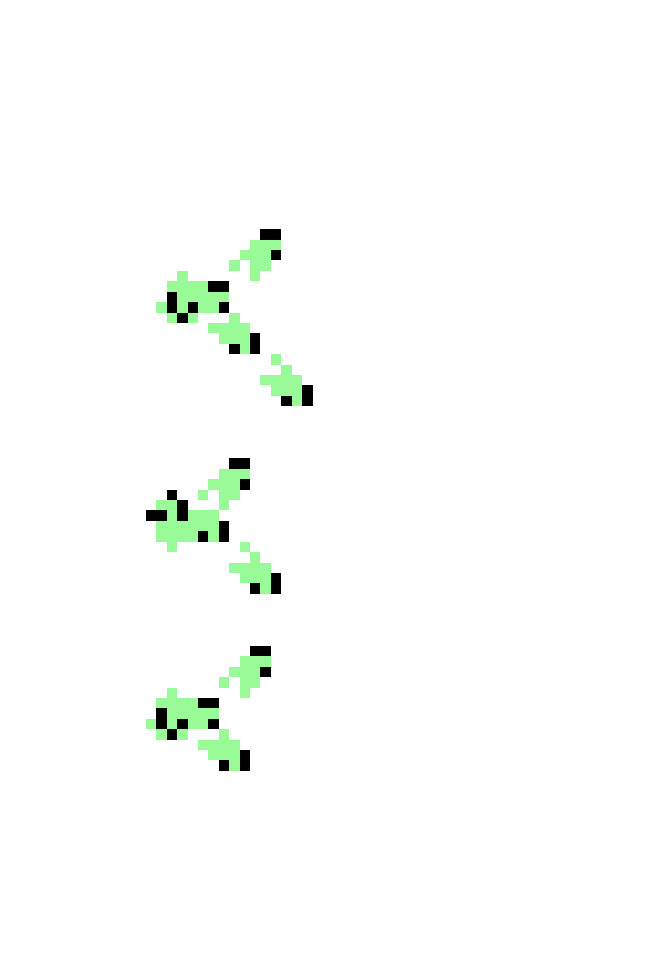}
\put(-10,5){(b)}
\end{overpic}
\end{minipage}
\end{center}
\vspace{-4ex}
\caption[glide+oscillator=glider-gun]
{\textsf{(a) three different collisions between a glider with an oscillator
create a glider-gun (b) shown after 43 time-steps.}}
\label{coll-3o}
\end{figure}

\begin{figure}[htb]
\begin{center}
\begin{minipage}[c]{.6\linewidth}
  \begin{overpic}[width=.3\linewidth, bb=145 65 239 162, clip=]{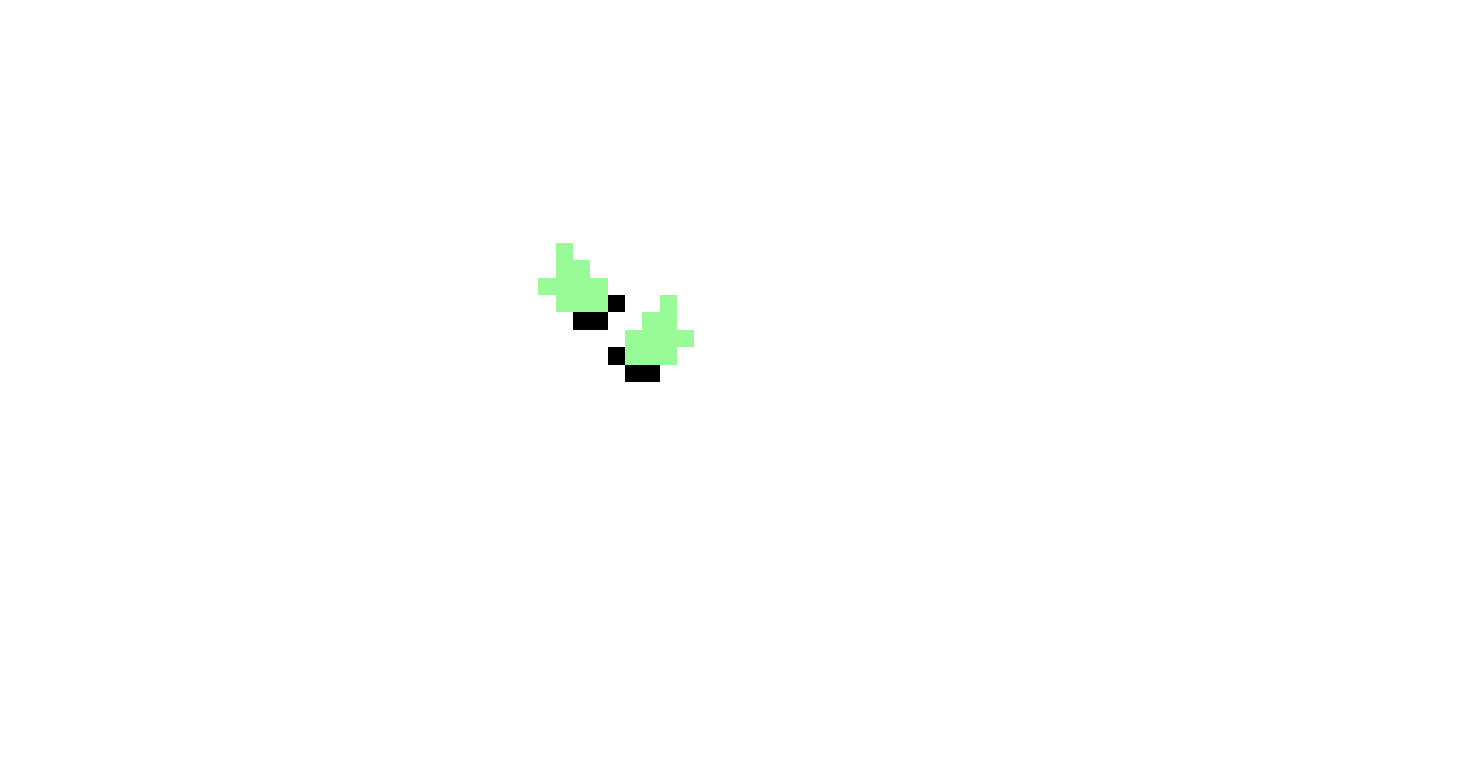}
\put(10,10){(a)}
\end{overpic}
\hfill
\begin{overpic}[width=.3\linewidth, bb=145 65 239 162, clip=]{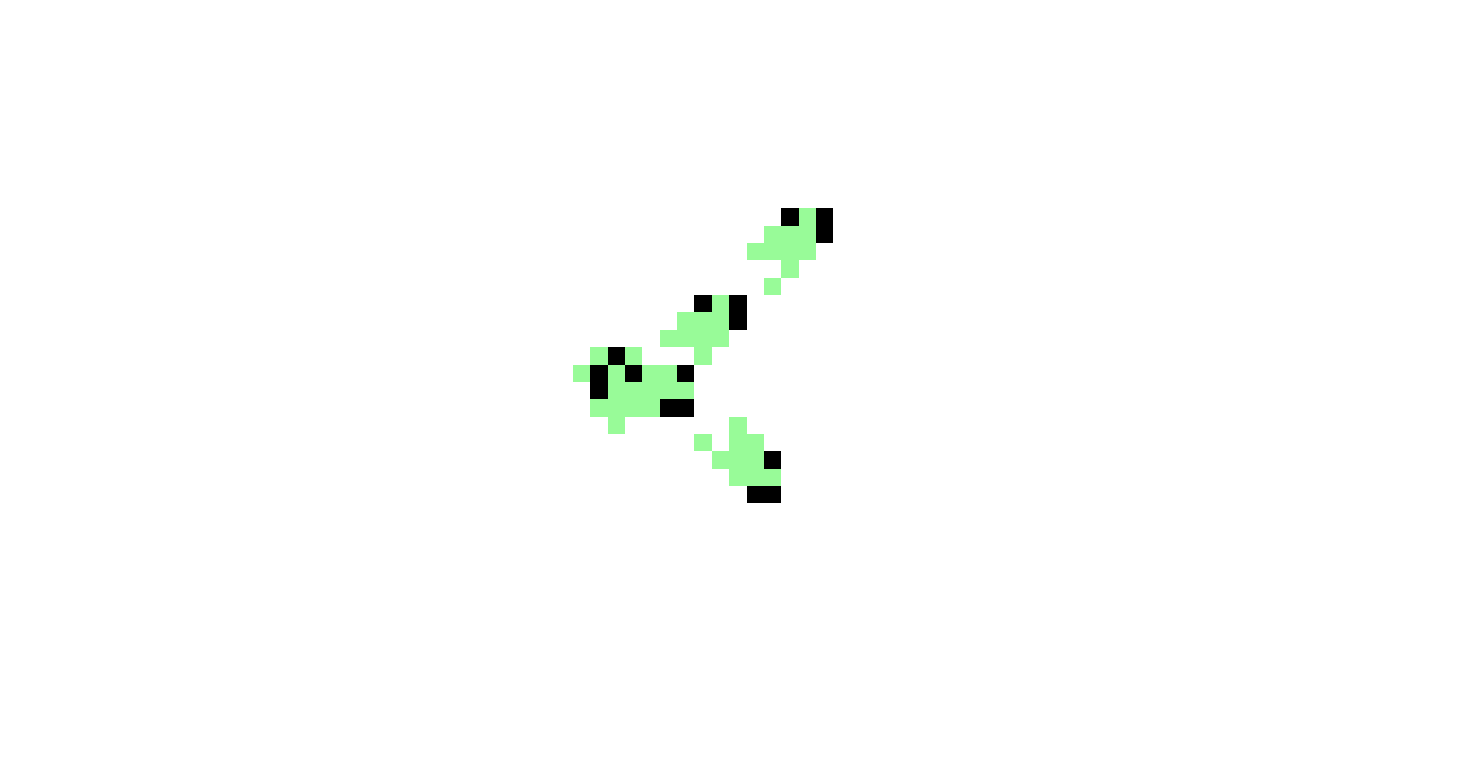}
\put(10,5){(b)}
\end{overpic}
\end{minipage}
\end{center}
\vspace{-4ex}
\caption[glide+glider=glider-gun]
{\textsf{(a) two gliders colliding at 90$^{\circ}$ create a glider-gun (b) shown after 48 time-steps.}}
\label{g+g-gg}
\end{figure}

Eaters that are able to stop a stream of gliders, are a necessary
component in the computation machinery. They can be derived from
still-lives or oscillators (figure~\ref{eater1}).
The glider-gun itself can be the outcome of a collision between a glider and an
oscillator (figures~\ref{coll-3o}),
or between two gliders (figure~\ref{g+g-gg}).\\[-3ex]

\clearpage

\begin{figure}[h] 
  \begin{overpic}[width=.32\textwidth, bb=171 165 340 316, clip=]{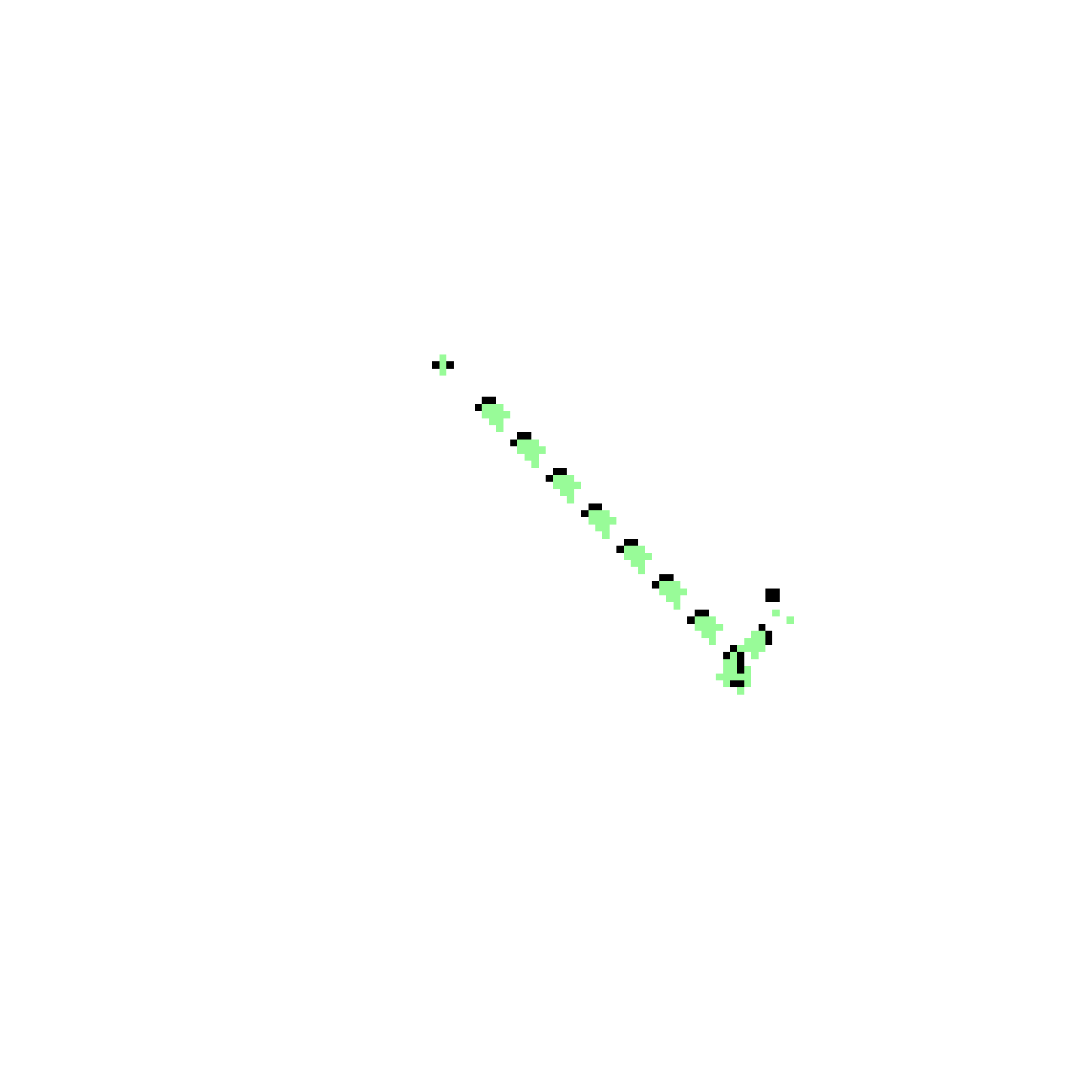}
\put(40,5){(a)}
\end{overpic}
\hfill
\begin{overpic}[width=.32\textwidth, bb=171 165 340 316, clip=]{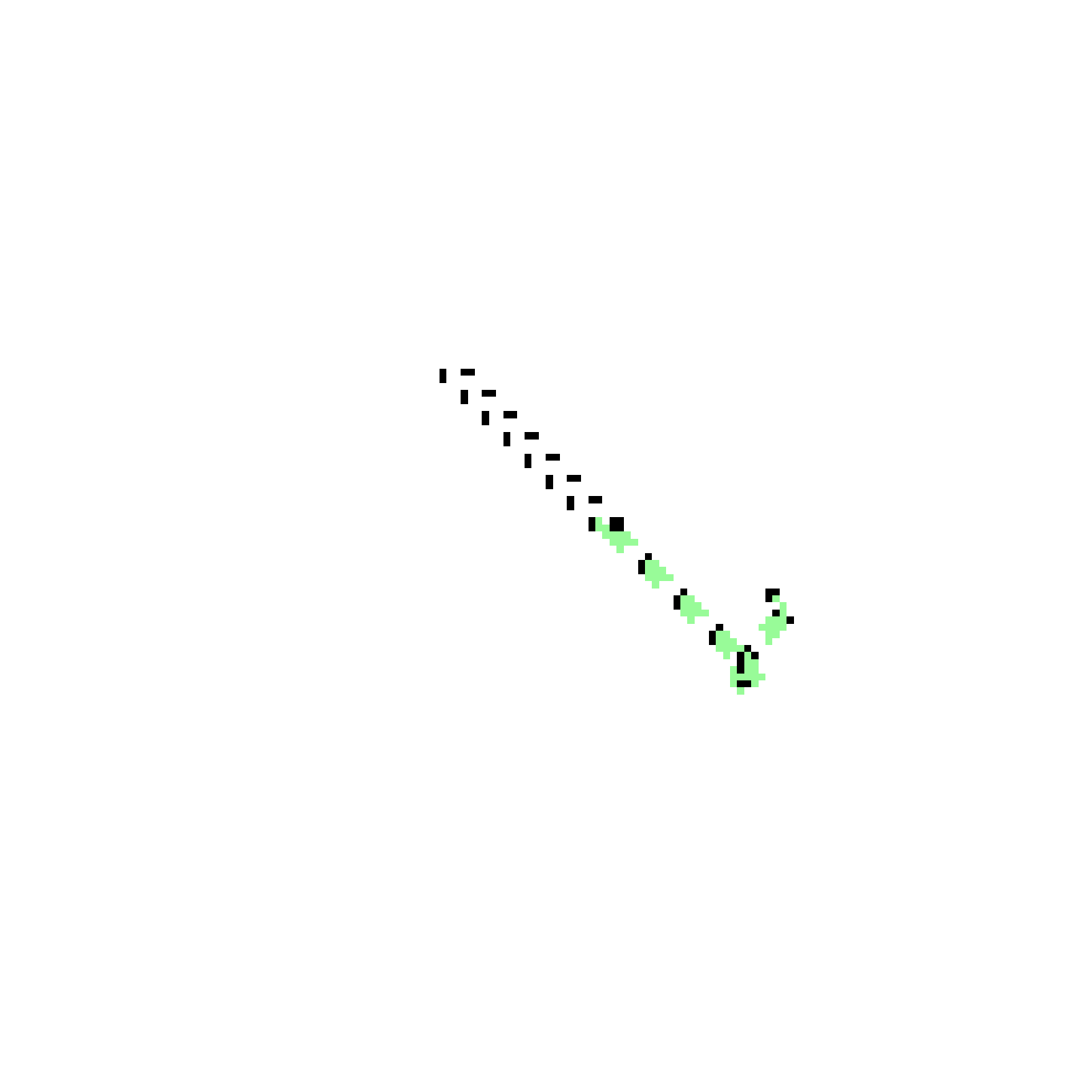}
\put(40,5){(b)}
\end{overpic}
\hfill
\begin{overpic}[width=.32\textwidth, bb=171 165 340 316, clip=]{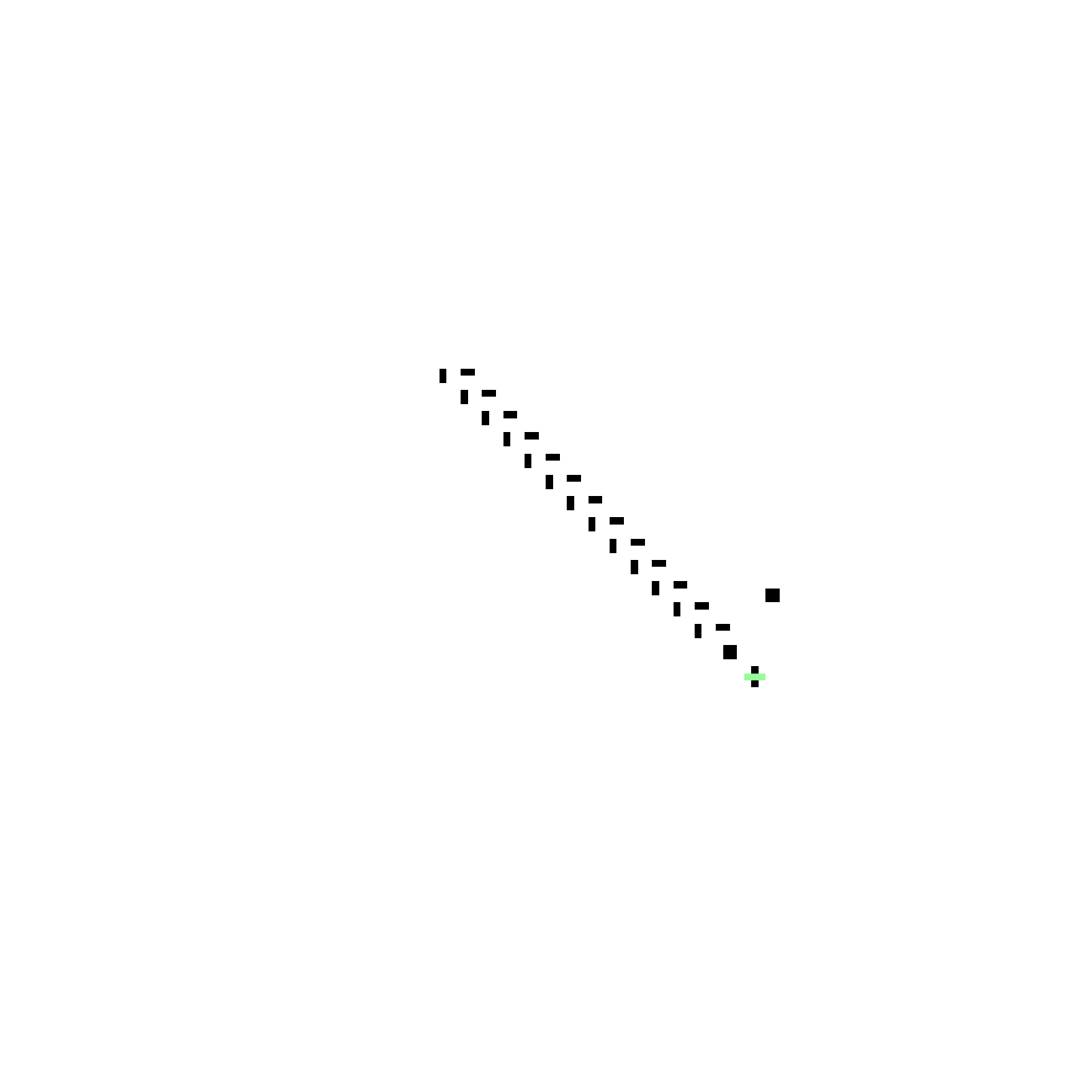}
\put(40,5){(c)}
\end{overpic}
\vspace{-2ex}
\caption{
{\textsf{Glider-gun stream (a) collides with an oscillator resuting in a retrograde stable
pattern (b) moving backwards that eventualy destroys the glider-gun (c).}}}
\label{pattern moving backwards}
\end{figure}

A particular but not infrequent collision situation can arise between a stream of gliders and
an oscillator which results in a retrograde stable pattern moving backwards, a sort of footprint.
This eventually destroys the originating glider-gun as illustrated in
figure~\ref{pattern moving backwards}.

\begin{figure}[htb]
\begin{center}
\begin{minipage}[c]{.4\linewidth}
  \begin{overpic}[width=.35\linewidth, bb=65 116 107 145, clip=]{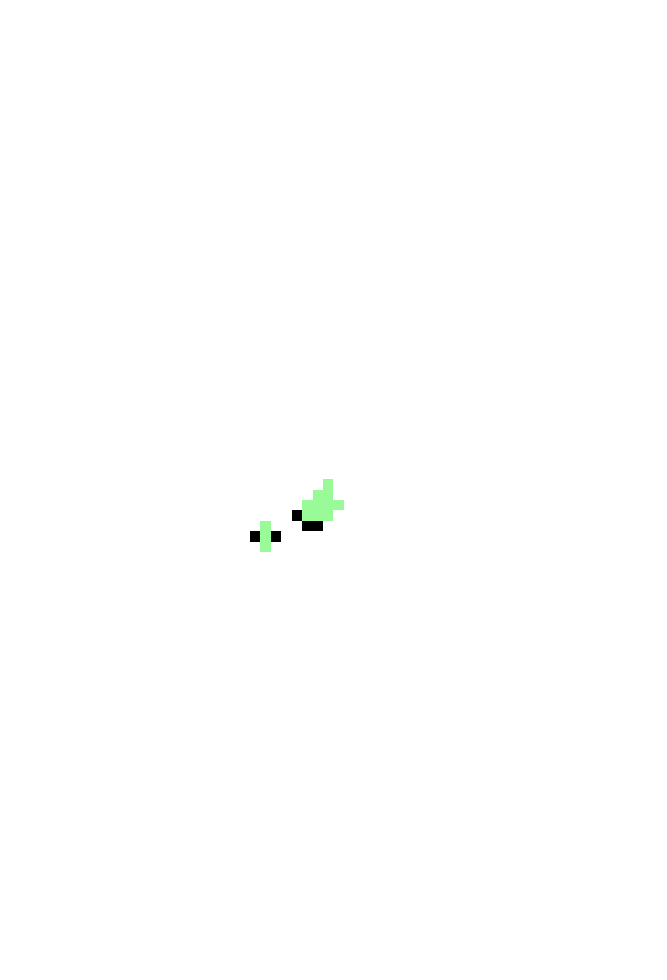}
\put(-10,5){(a)}
\end{overpic}
\hfill
\begin{overpic}[width=.35\linewidth, bb=65 116 107 145, clip=]{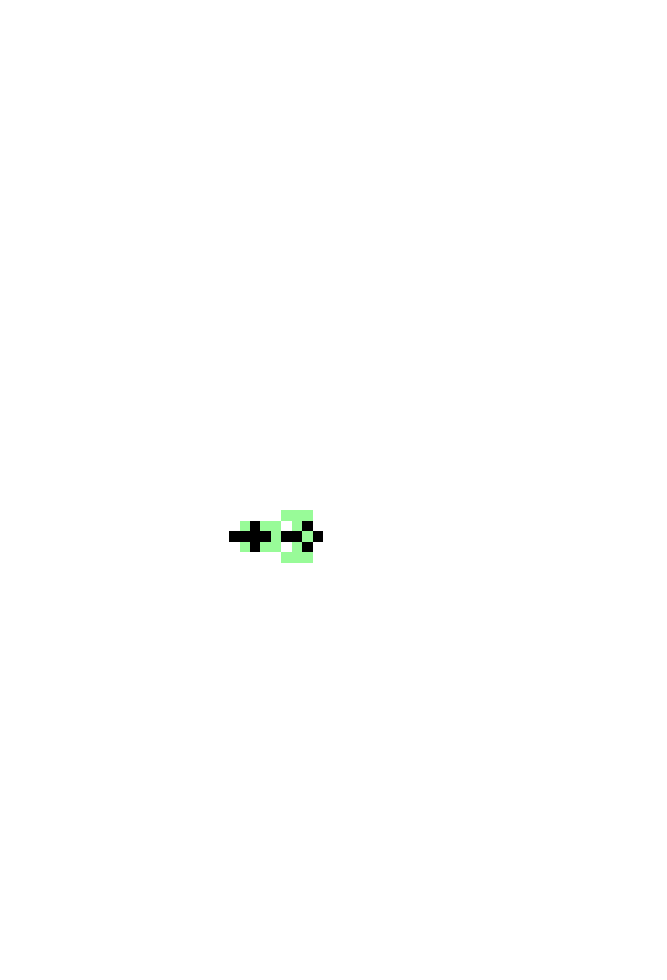}
\put(-15,5){(b)}
\end{overpic}
\end{minipage}\\
\begin{minipage}[c]{1\linewidth}
\includegraphics[width=.073\linewidth, bb=60 109 97 144, clip=]{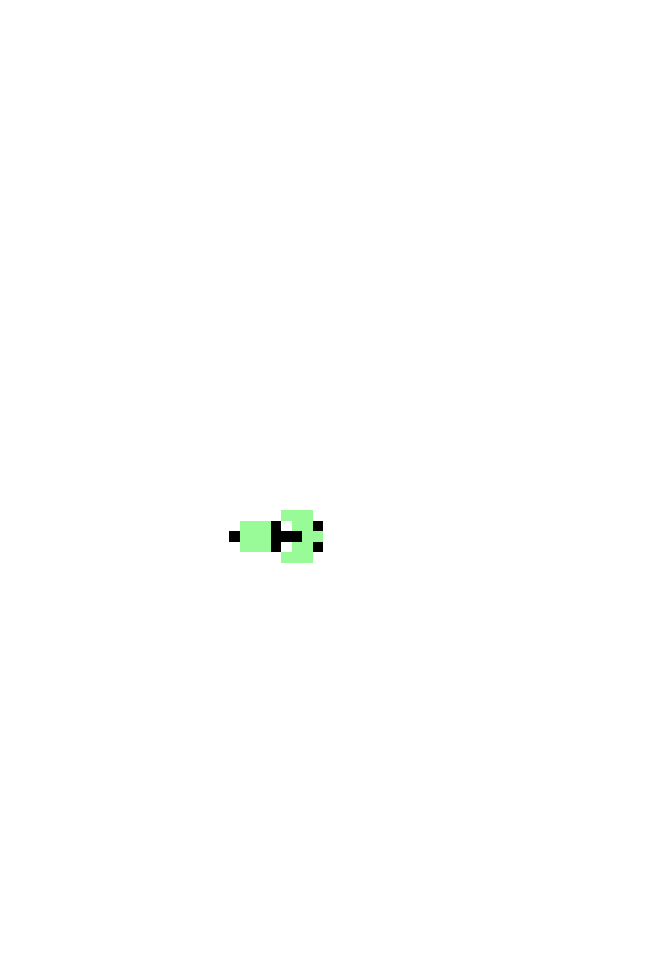} \hfill
\includegraphics[width=.073\linewidth, bb=60 109 97 144, clip=]{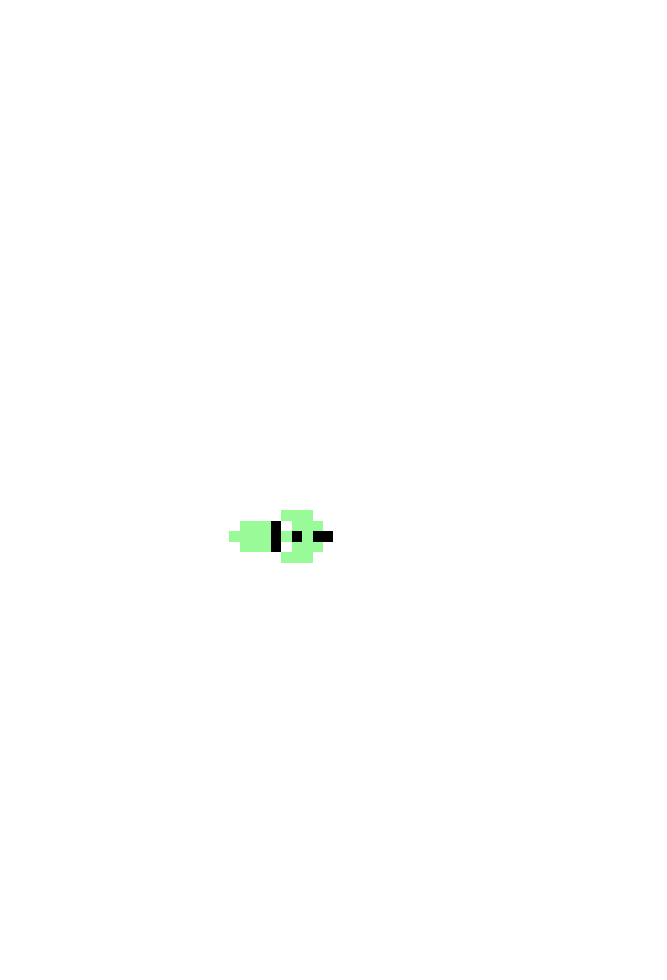} \hfill
\includegraphics[width=.073\linewidth, bb=60 109 97 144, clip=]{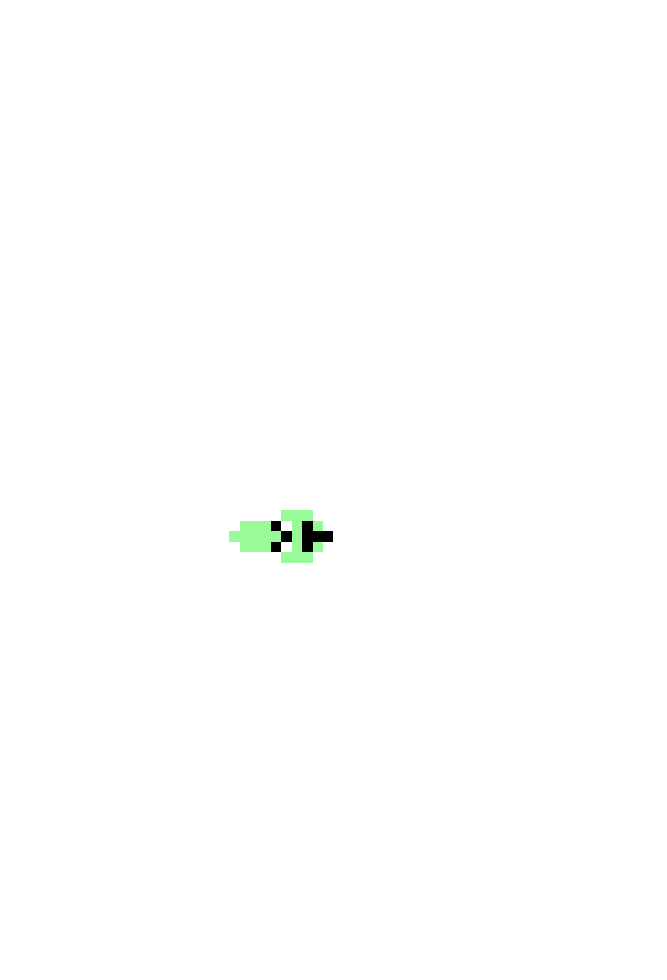} \hfill
\includegraphics[width=.073\linewidth, bb=60 109 97 144, clip=]{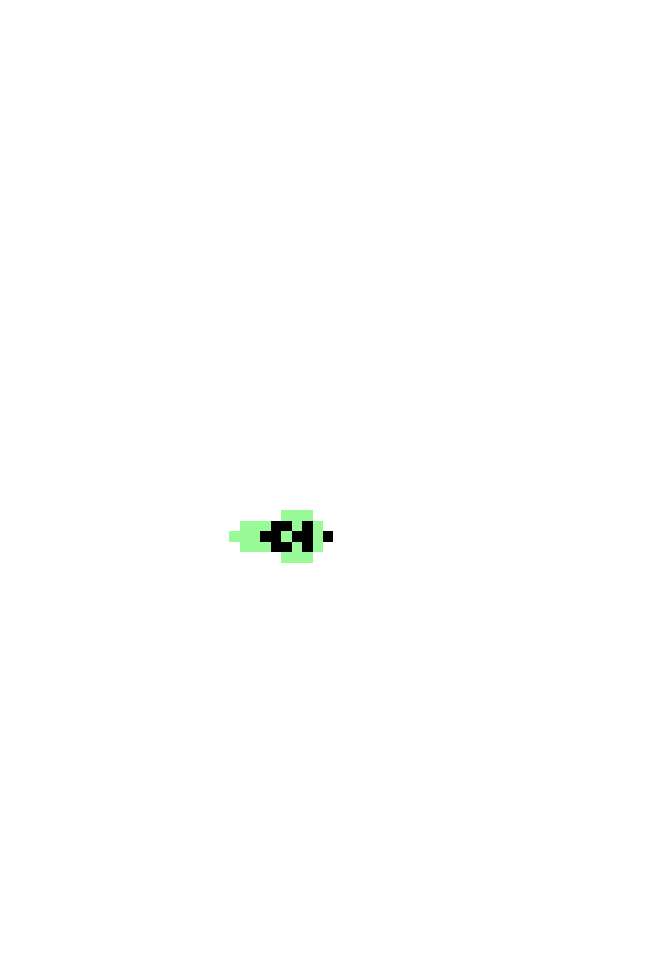} \hfill
\includegraphics[width=.073\linewidth, bb=60 109 97 144, clip=]{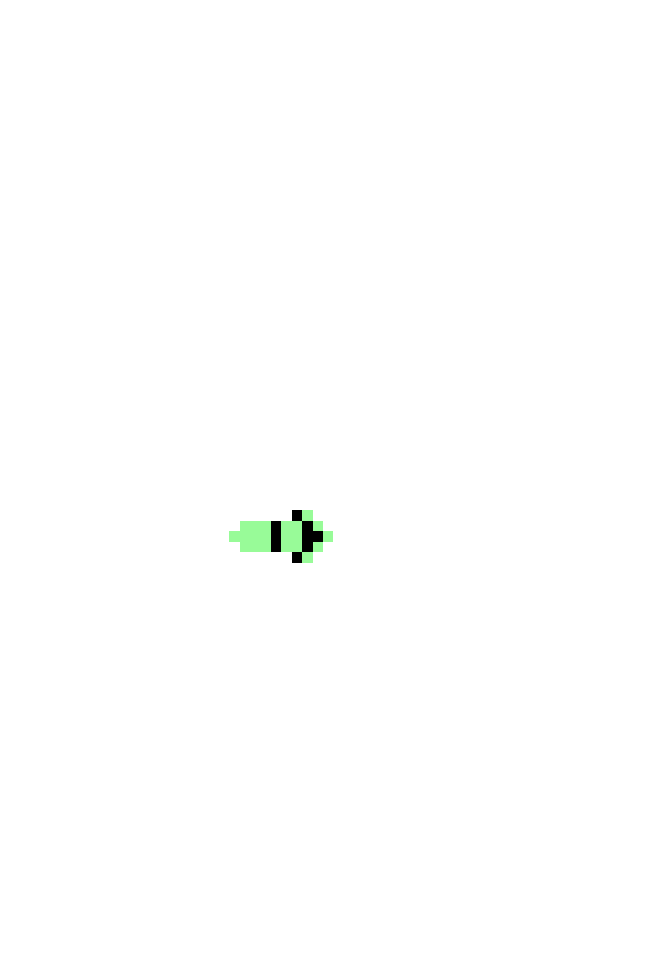} \hfill
\includegraphics[width=.073\linewidth, bb=60 109 97 144, clip=]{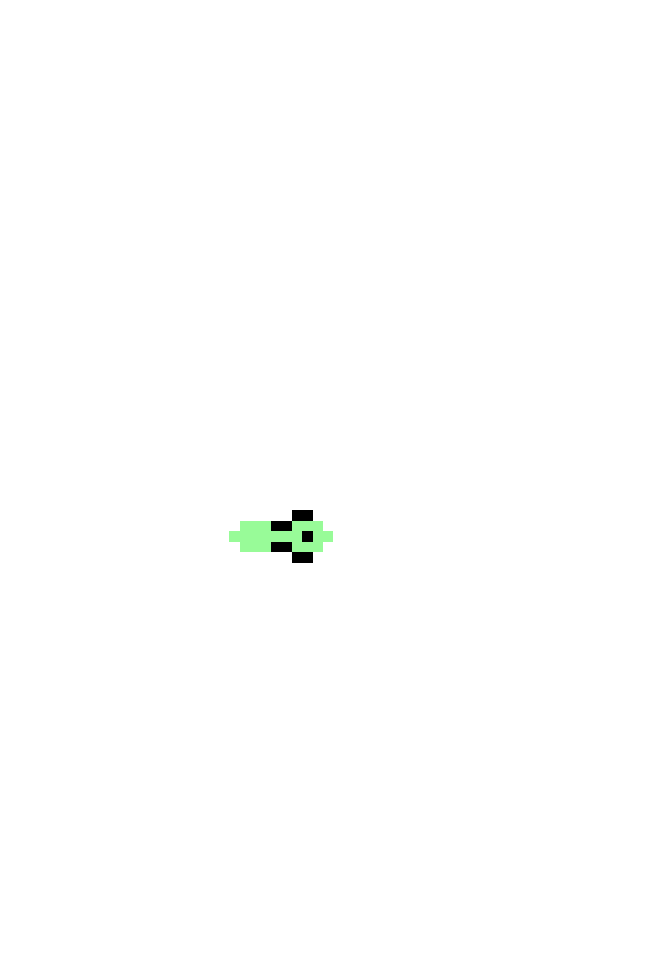} \hfill
\includegraphics[width=.073\linewidth, bb=60 109 97 144, clip=]{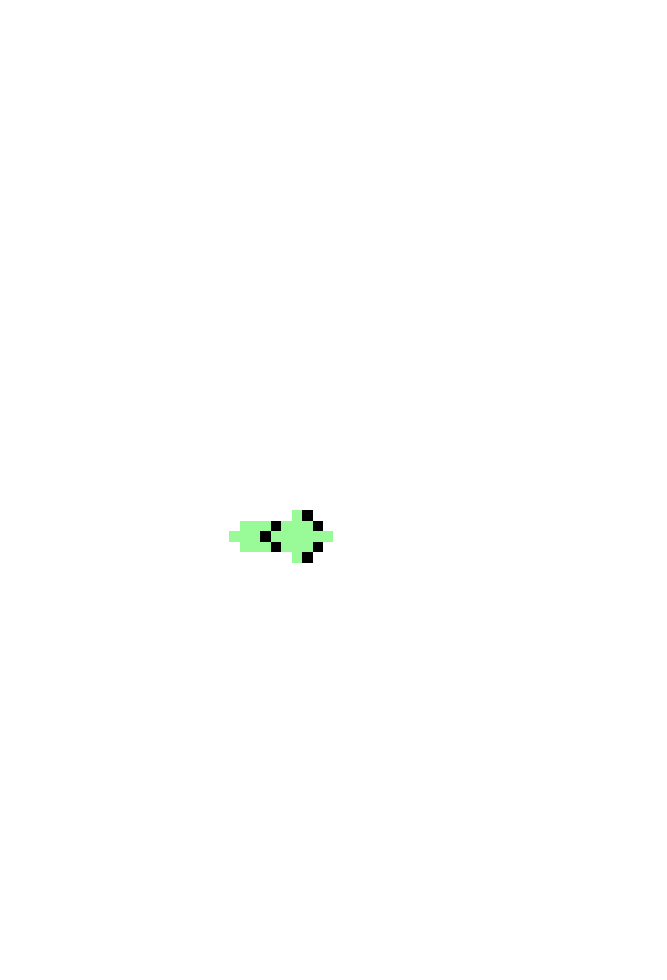} \hfill
\includegraphics[width=.073\linewidth, bb=60 109 97 144, clip=]{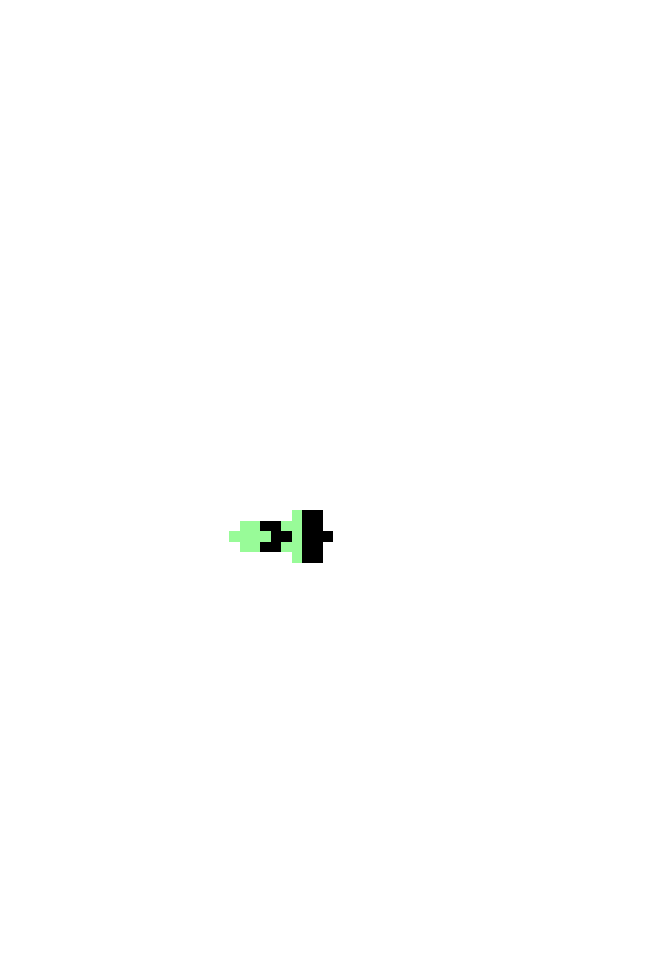} \hfill
\includegraphics[width=.073\linewidth, bb=60 109 97 144, clip=]{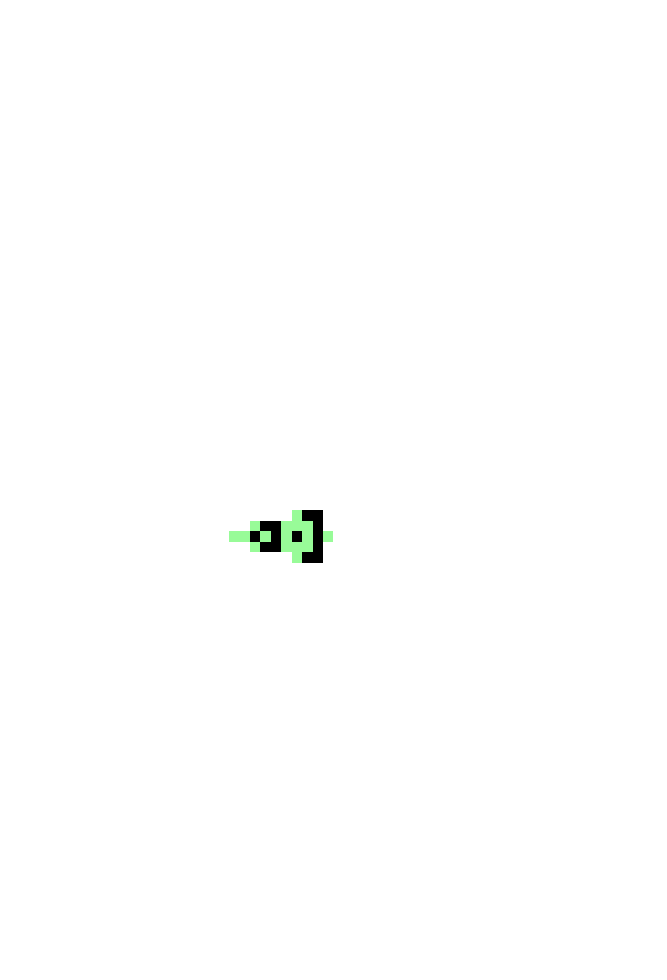} \hfill
\includegraphics[width=.073\linewidth, bb=60 109 97 144, clip=]{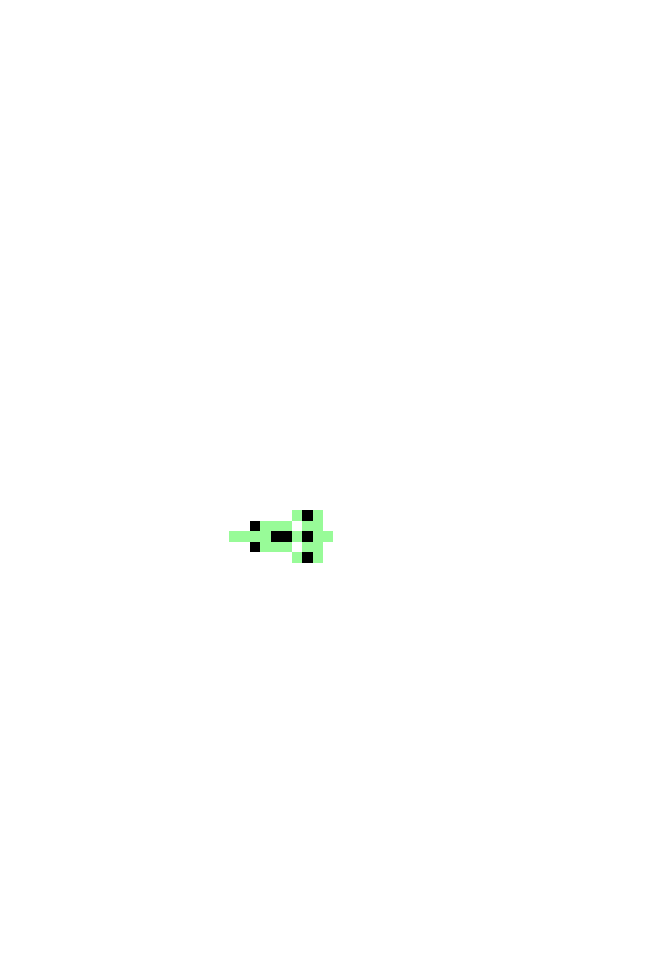} \hfill
\includegraphics[width=.073\linewidth, bb=60 109 97 144, clip=]{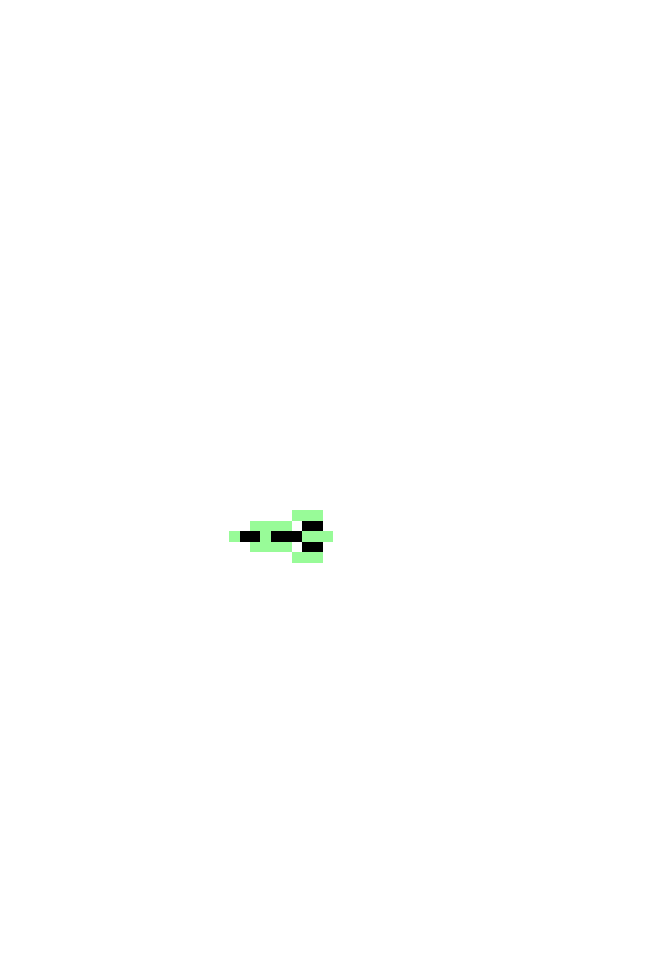} \hfill
\includegraphics[width=.073\linewidth, bb=60 109 97 144, clip=]{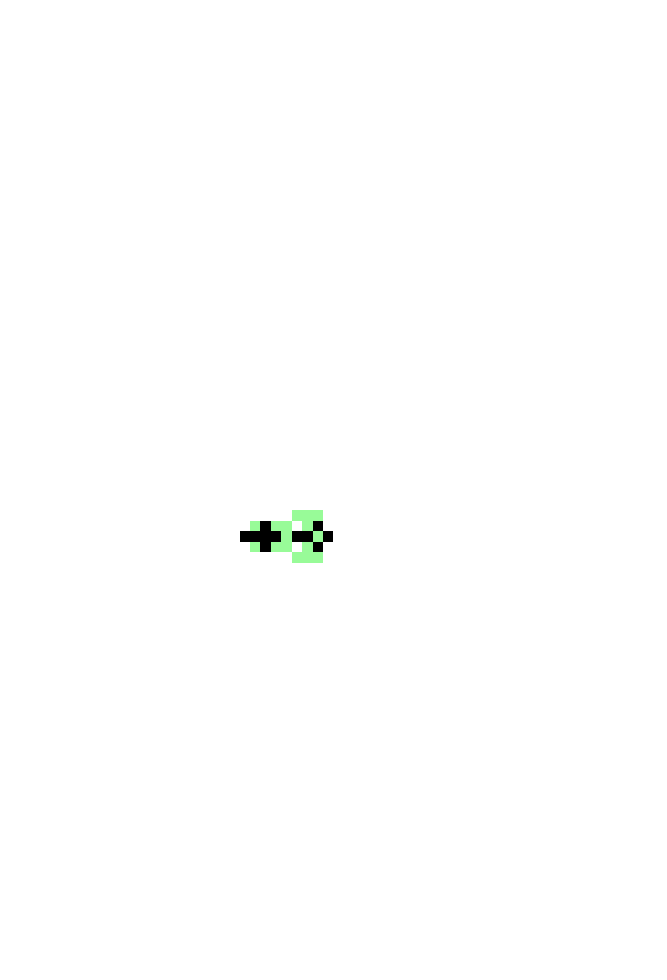} \hfill
\end{minipage}
\end{center}
\vspace{-4ex}
\caption[glide+oscillator=space-ship]
{\textsf{(a) a glider collides with an oscillator creating a slow moving 
space-ship (b) shown after 25 time-steps. The 12 phases of the space-ship are shown.}}
\label{coll-ng}
\vspace{-4ex}
\end{figure}

\begin{figure}[h]
\begin{center}
  \includegraphics[width=1\linewidth, bb=33 8 369 89, clip=]{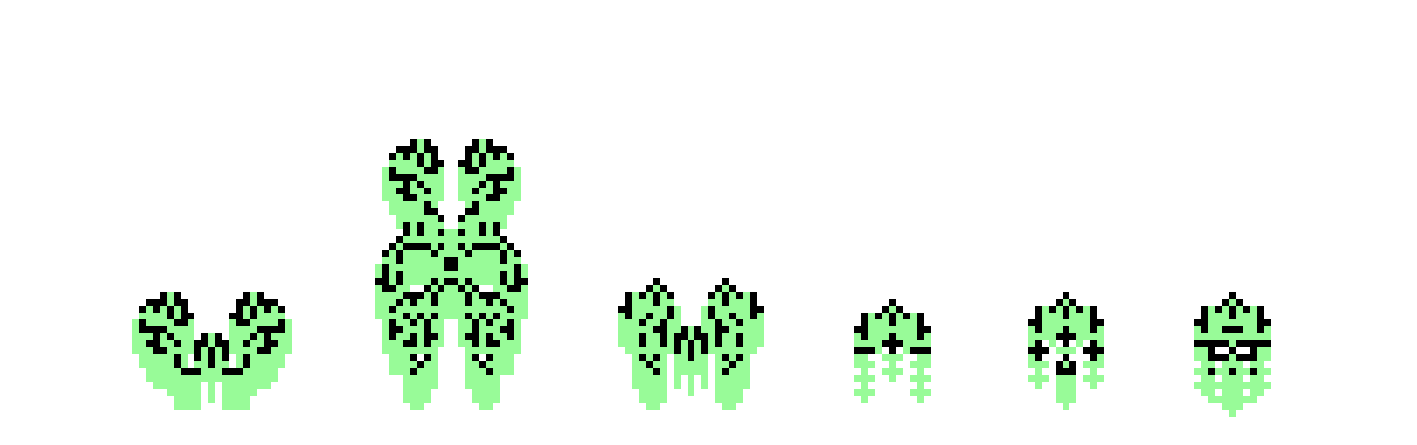} 
\end{center}
\vspace{-5ex}
\caption{ 
{\textsf{Six large space-ships moving North with speed $c$/2. Periods, from
left to right, are 2, 2, 2, 4, 4, 4.}}} 
\label{sships}
\end{figure}

A small slow moving space-ship (an orthogonal glider) can result from a collision between
a glider and an oscillator, as shown in figure~\ref{coll-ng}. The spaceship that emerges
has a frequency of 12 and speed of $c$/12, so it takes 12 time-steps to advance one cell.
Larger space-ships with various frequencies are shown in figure~\ref{sships}.\\[-6ex]

\enlargethispage{6ex}
\clearpage

A compound glider-gun (GG2) can be built from two interlocking GG1
glider-guns. GG2 shoots two glider streams in opposite directions with
a frequency of 40 time-steps and a glider spacing is 10 cells (twice
GG1). The dynamics depend on glider streams colliding at 90$^{\circ}$
resulting in the destruction of one glider-stream, and alternate
gliders in the other glider-stream. Collisions leave behind a
sacrificial ``eater'' which destroys one of the next pair of incoming
gliders.

Two GG2 glider-guns can be combined into a larger compound glider-gun (GG4, figure~\ref{GG4})
where analogous collisions result in doubling the GG2
frequency and spacing, so the GG4 glider-stream has a frequency of 80 time-steps and
spacing of 20 cells. This doubling of glider-stream frequency and spacing with greater compound
glider-guns can be continued without limit.

\section{Logical Universality and Logical Gates}
\label{Logical Universality and Logical Gates}

Post's Functional Completeness Theorem\cite{Post41,Francis90}
established that it is possible to make a disjuntive (or conjuctive)
normal form formula using the logical gates NOT, AND and OR.
Conway applies this as his 3rd condition for a cellular automata to be
universal in the full sense. The three conditions, applied to the
game-of-Life\cite{Berlekamp1982}, state
that the system must be capable of the following:

\begin{s-enumerate}
\item Data storage or memory.
\item Data transmission requiring wires and an internal clock.
\item Data processing requiring a universal set of logic gates NOT, AND, and OR, to satisfy
  negation, conjunction and disjunction.
\end{s-enumerate} 

This section is confined to demonstrating the logical gates, so Conway's
condition~3, for universality in the logical sense.
To demonstrate universality in Conway's full sense\footnote{Alternatively, 
full universality could be proved in terms of the Turing Machine, as was done by
Randall\cite{Randall2002}.}  it would be necessary to also prove
conditions 1 and 2.

\begin{figure}[htb]
\begin{center}
\begin{minipage}[c]{.6\linewidth}
\includegraphics[width=.17\linewidth, bb=65 65 118 135, clip=]{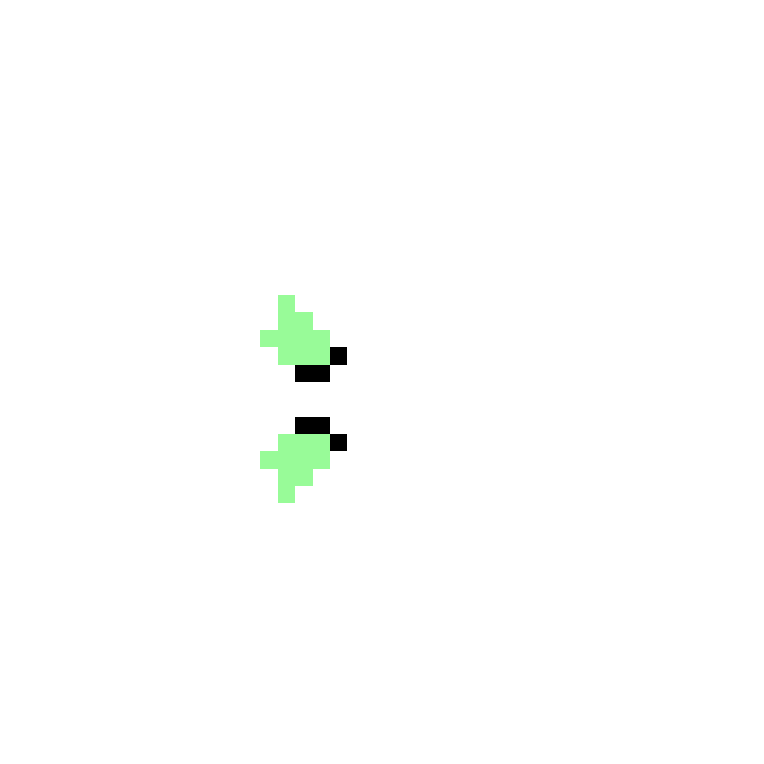}
\hfill
\includegraphics[width=.17\linewidth, bb=65 65 118 135, clip=]{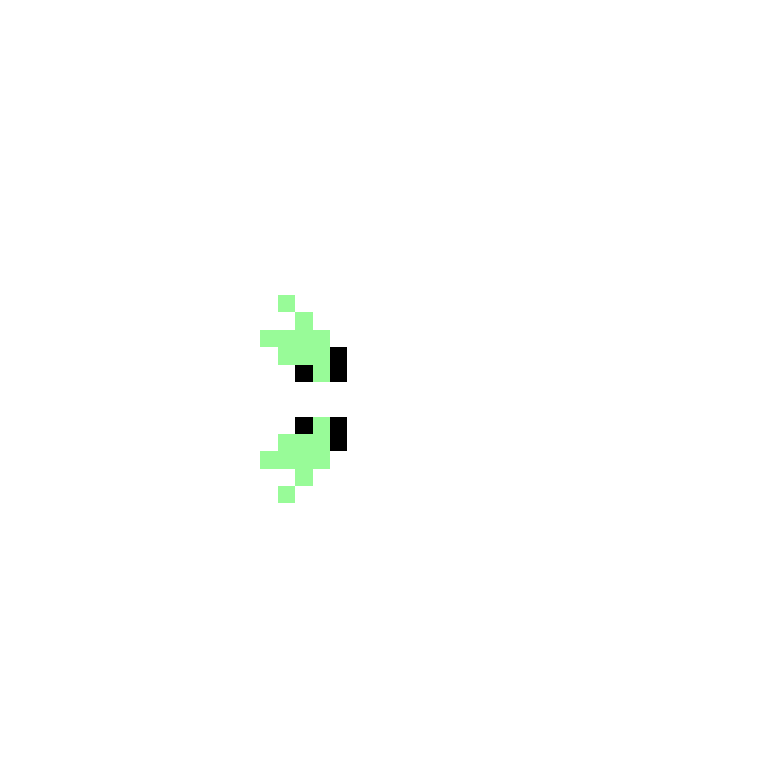}
\hfill
\includegraphics[width=.17\linewidth, bb=65 65 118 135, clip=]{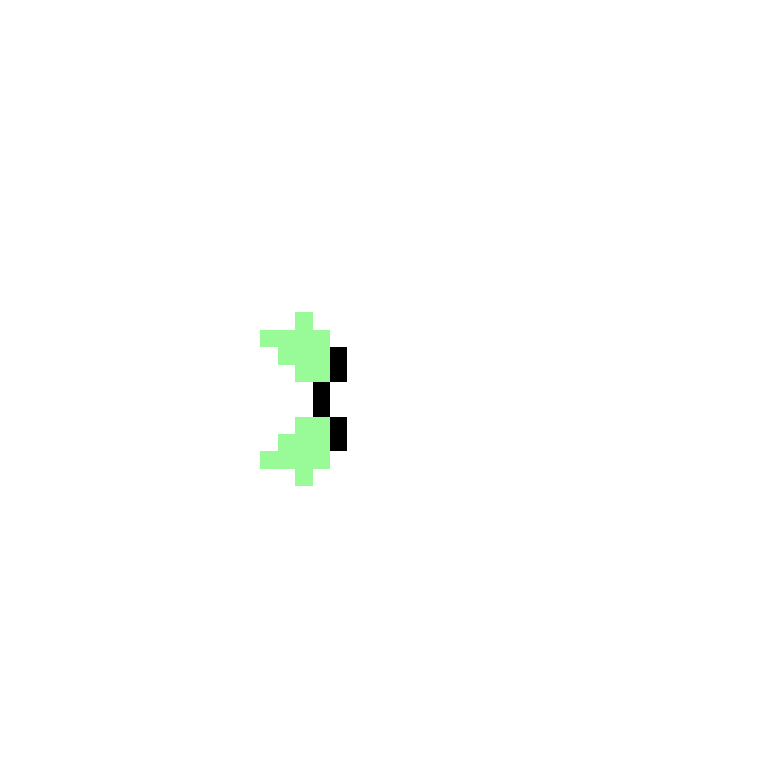}
\hfill
\includegraphics[width=.17\linewidth, bb=65 65 118 135, clip=]{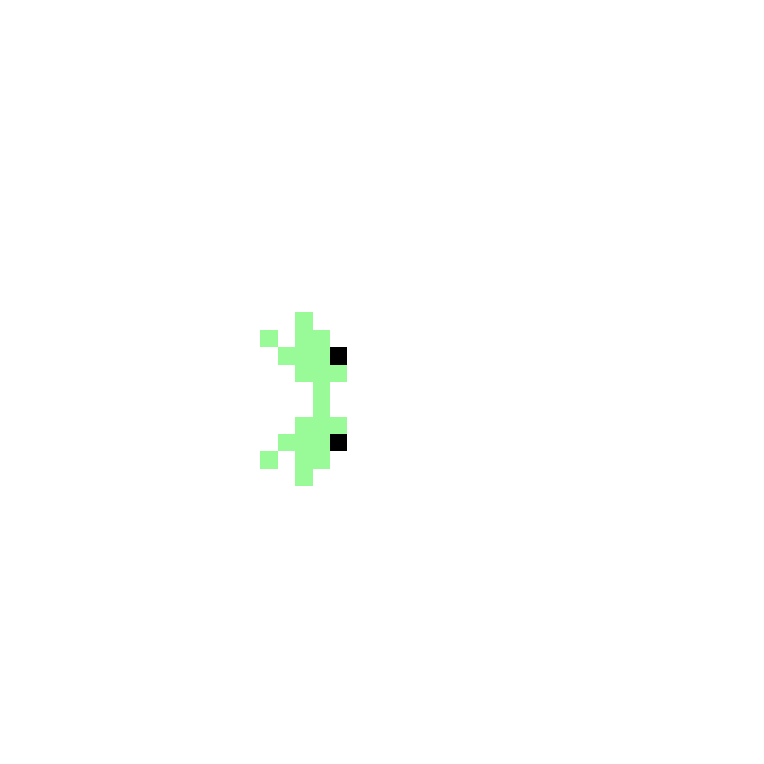}
\hfill
\includegraphics[width=.17\linewidth, bb=65 65 118 135, clip=]{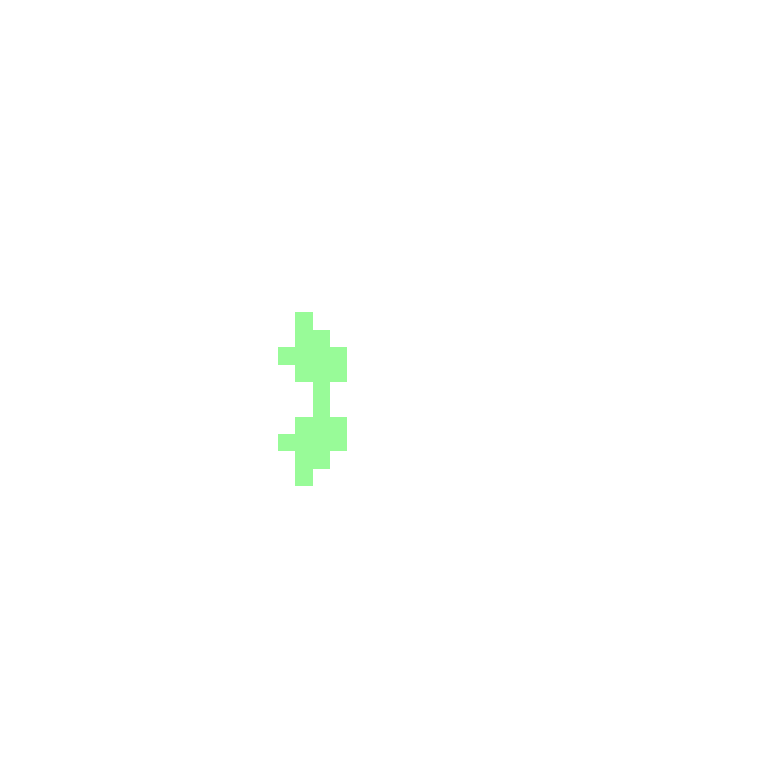}
\end{minipage}
\end{center}
\vspace{-4ex}
\caption[Two gliders collide self-destruct]
{\textsf{Two gliders colliding at 90$^{\circ}$ self-destruct. 5 consecutive
time-steps are shown. This is a key collision in making logical gates.
Head-on collisions also self destruct, but are not as useful in this context.}}
\label{gcol1}
\end{figure}

We propose that the basic existential ingredients for constructing
logical gates, and thus logical universality, are as follows:

\begin{s-enumerate}

\item A glider-gun or ``pulse generator'', 
that sends a stream of gilders\footnote{Gliders are not listed separately because
they are implicit in the glider-gun.} into space 
(figures~\ref{glider-gun GG1} and \ref{cycle-state}).

\item An eater, based on a still-life or oscillator, that destroys an incoming glider and
survives the collision, so can stop a glider stream (figure~\ref{eater1}).

\item Complete self-destruction when two gliders collide at an angle. 
Any debris must quickly dissipate, and the gap between gliders must be sufficient
so as not to interfere with the next glider collision 
(figure~\ref{gcol1}).

\end{s-enumerate}

\begin{figure}[h]
\begin{center}
\begin{minipage}[c]{.8\linewidth} 
  \fbox{\includegraphics[height=.5\linewidth, bb=5 9 176 271, clip=]{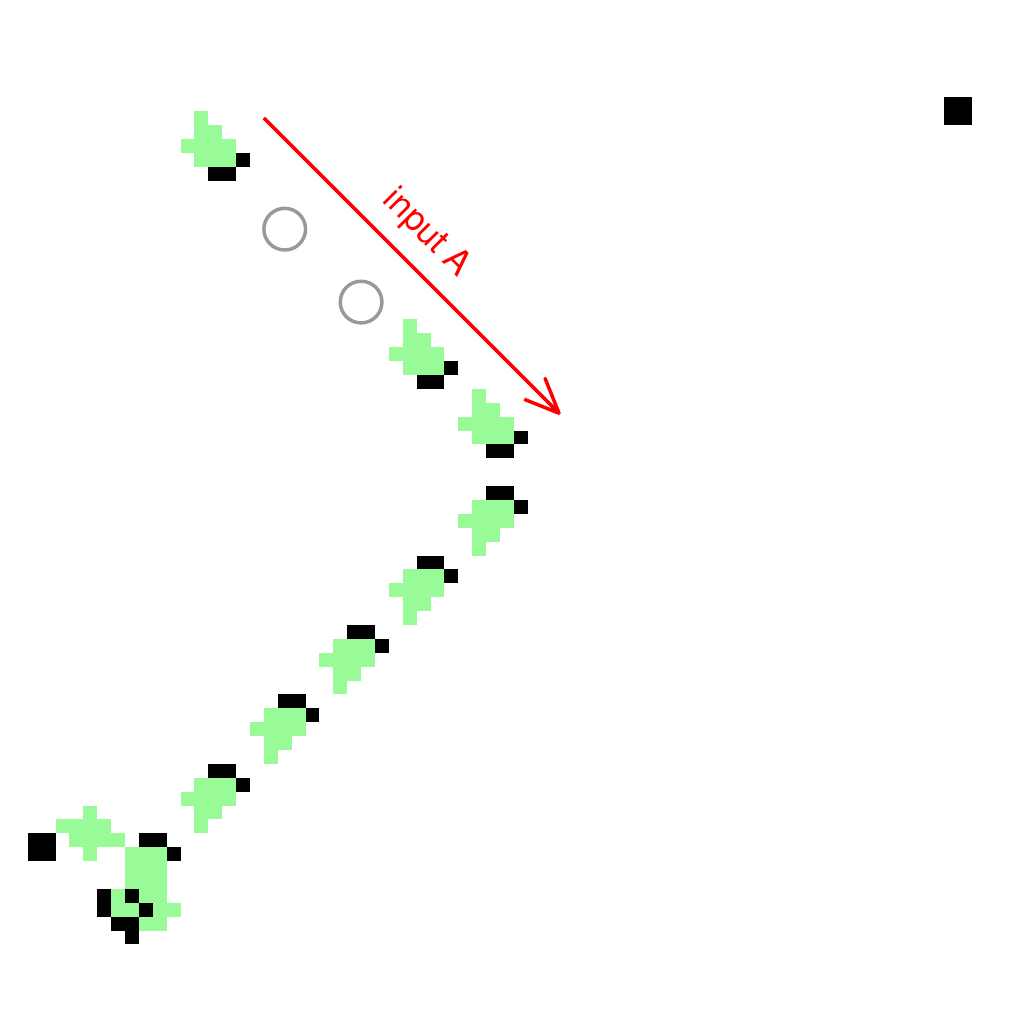}}
\hfill
 \fbox{\includegraphics[height=.5\linewidth, bb=5 9 287 271, clip=]{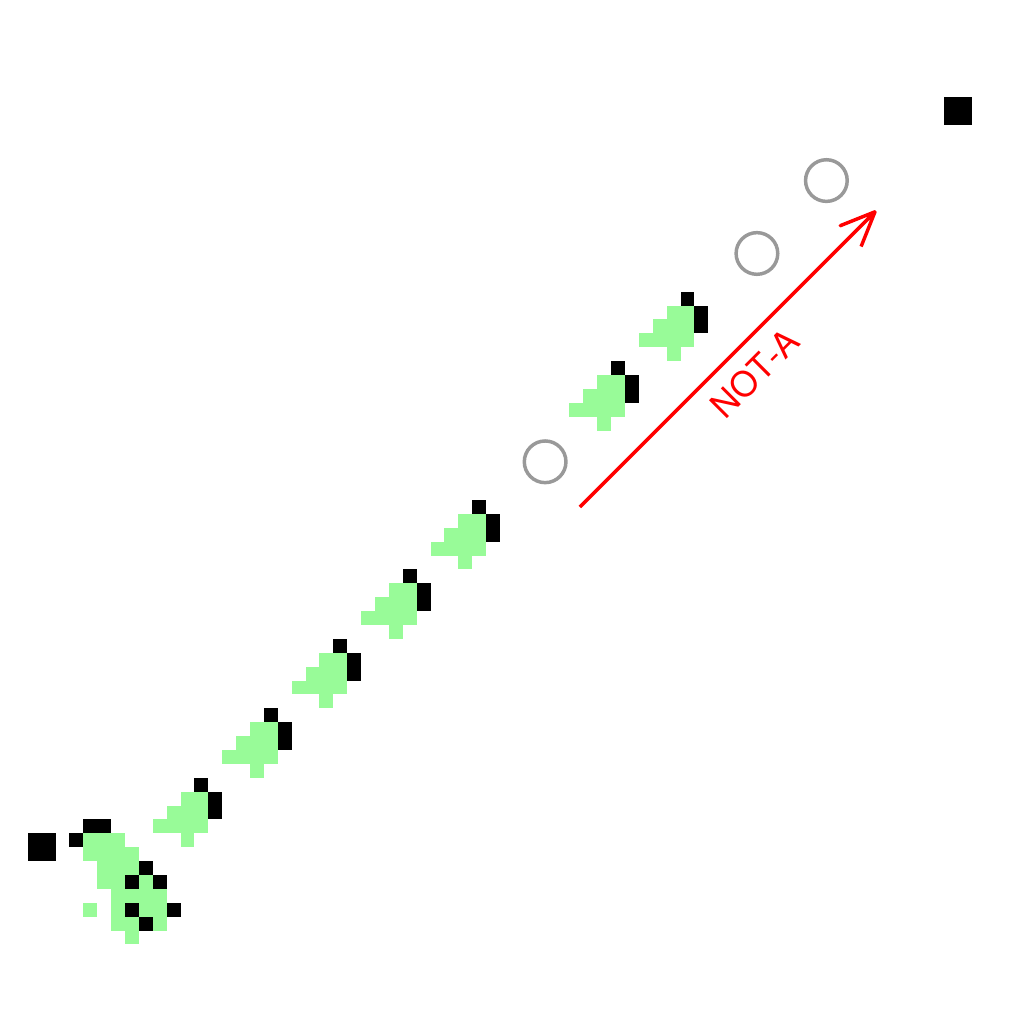}}
\end{minipage}
\end{center}
\vspace{-3ex}
\caption[NOT gate]
{\textsf{An example of the NOT gate: ($\neg 1, 1 \rightarrow$ 0 and $0
    \rightarrow$ 1) or inverter, which transforms a stream of data to
    its compliment, represented by gliders and gaps. 
The 5-bit input string A (11001) moving SE interacts 
with a GG1 glider-stream moving NE, resulting in NOT-A (00110) moving
NE, shown after 94 time-steps.
}}
\label{snot1A+}
\end{figure}

\begin{figure}[htb]
\begin{center}
\begin{minipage}[c]{.87\linewidth} 
  \fbox{\includegraphics[height=.65\linewidth, bb=74 77 200 361, clip=]{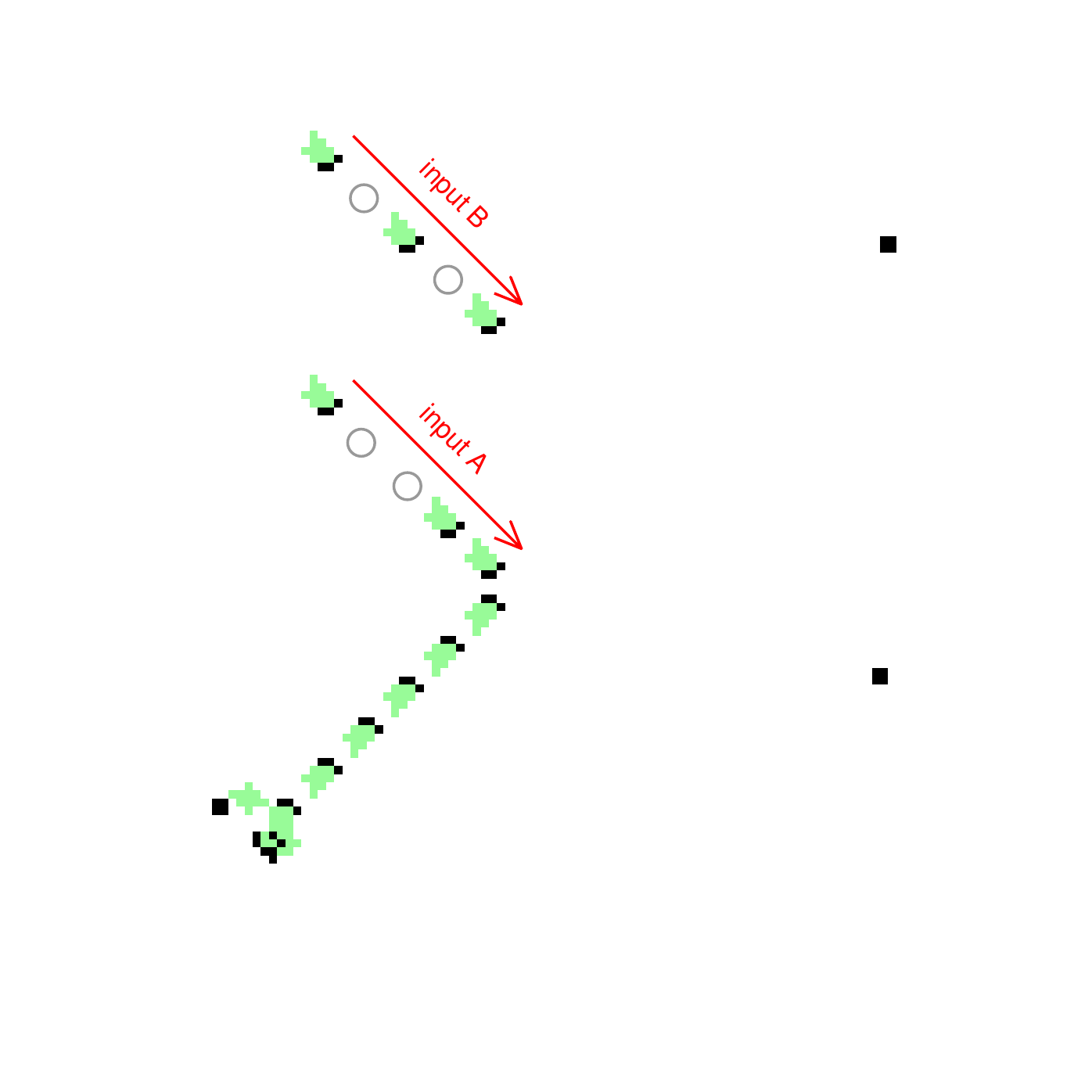}}
\hfill
  \fbox{\includegraphics[height=.65\linewidth, bb=74 77 334 361, clip=]{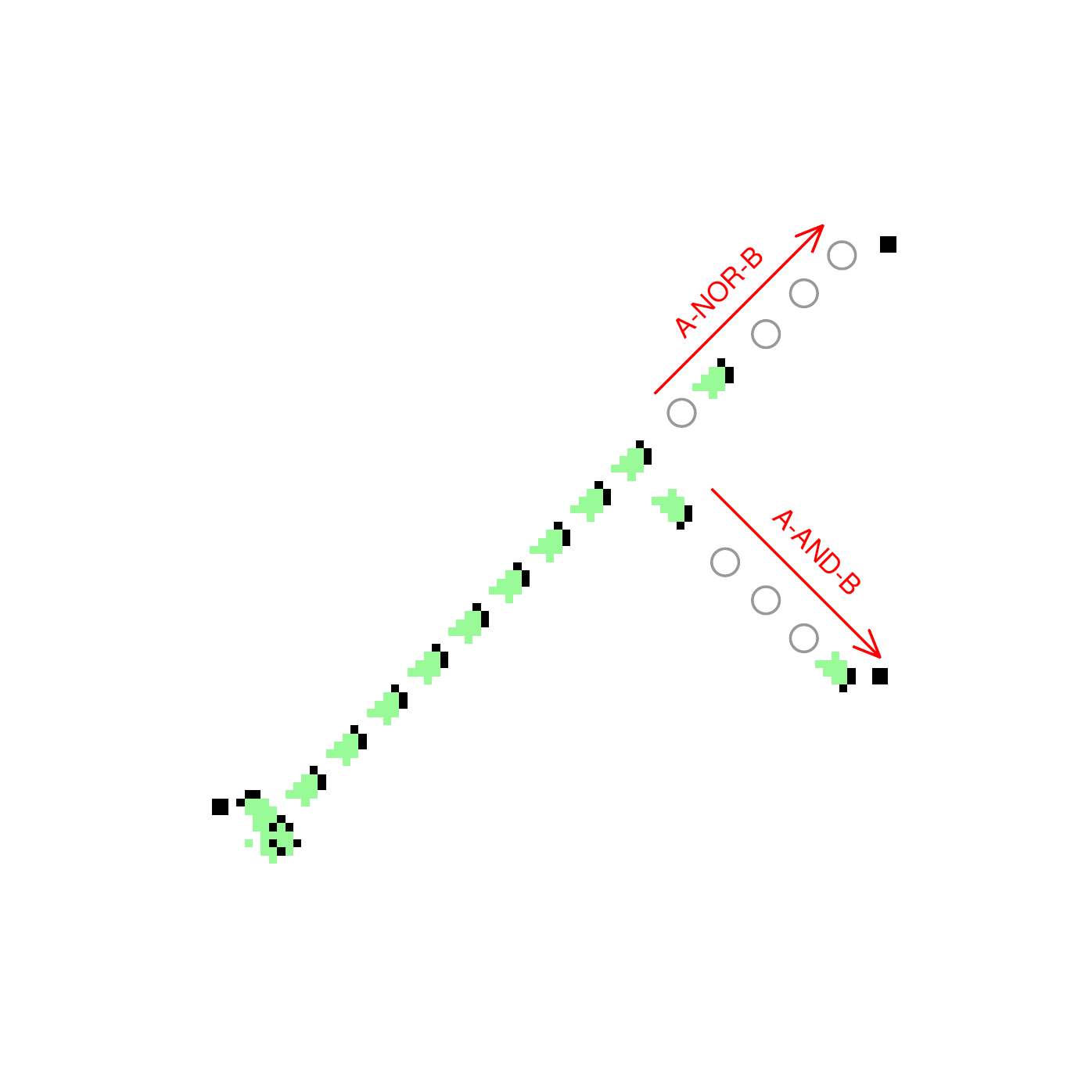}}
\end{minipage}
\end{center}
\vspace{-3ex}
\caption[AND gate]
{\textsf{An example of the AND gate (1 $\wedge$ 1 $\rightarrow$ 1,
    else $\rightarrow$ 0) making a conjunction between two streams of
    data, represented by gliders and gaps.
The 5-bit input strings A (11001) and B (10101) both moving SE
interact  with a GG1 glider-stream moving NE, resulting
in \mbox{A-AND-B} (10001) moving SE shown after 174 time-steps. The dynamics
making this AND gate first makes an intermediate \mbox{NOT-A} string
00110 (as in figure~\ref{snot1A+}) which then interacts with input string
B to simultaneously produce both the A-AND-B string moving SE
described above, and also the \mbox{A-NOR-B} string 00010 moving NE.}}
\label{sand1A+}
\end{figure}

\begin{figure}[htb]
\begin{center}
\begin{minipage}[c]{1\linewidth} 
  \fbox{\includegraphics[height=.8\linewidth, bb=29 33 163 430, clip=]{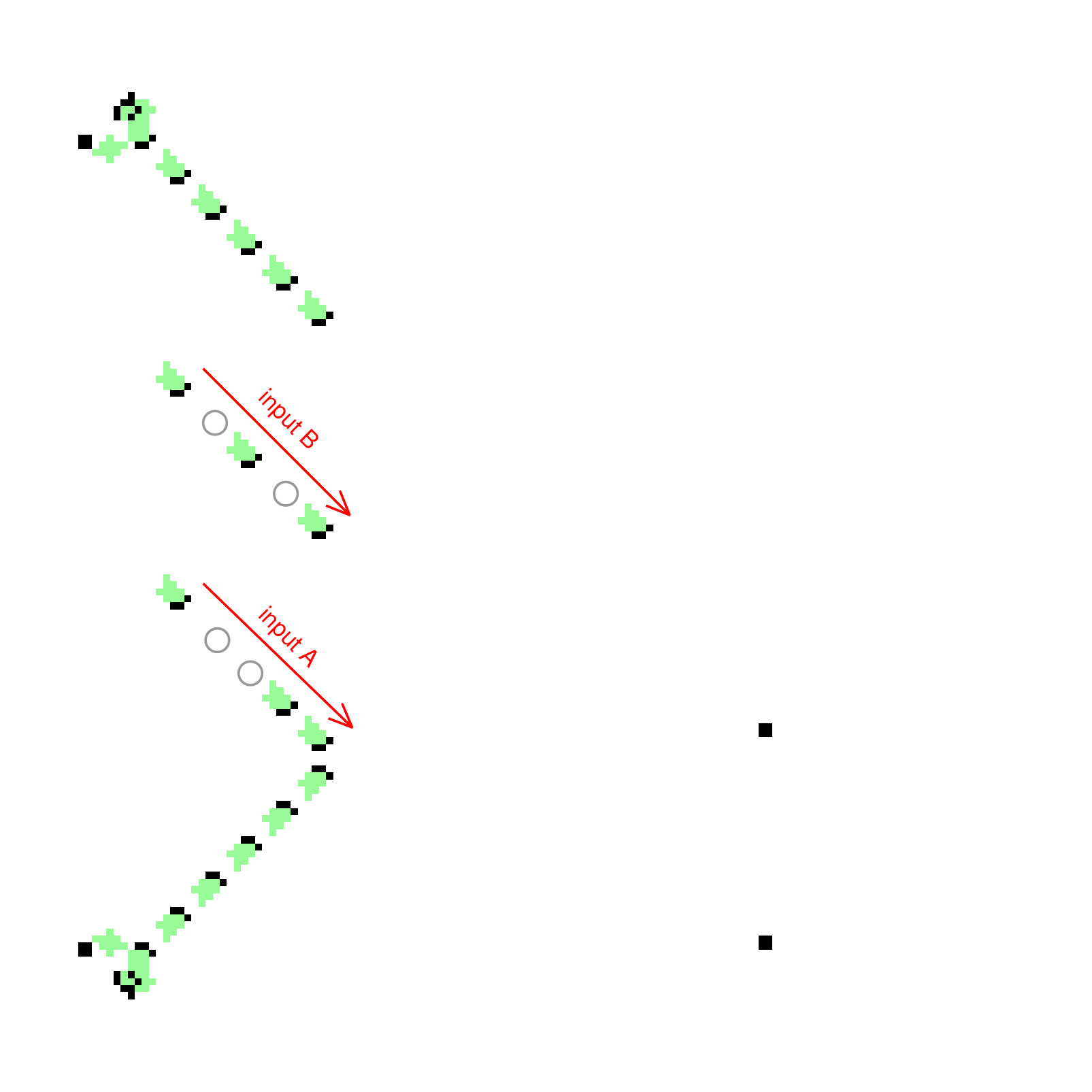}}
\hfill
  \fbox{\includegraphics[height=.8\linewidth, bb=29 33 337 430, clip=]{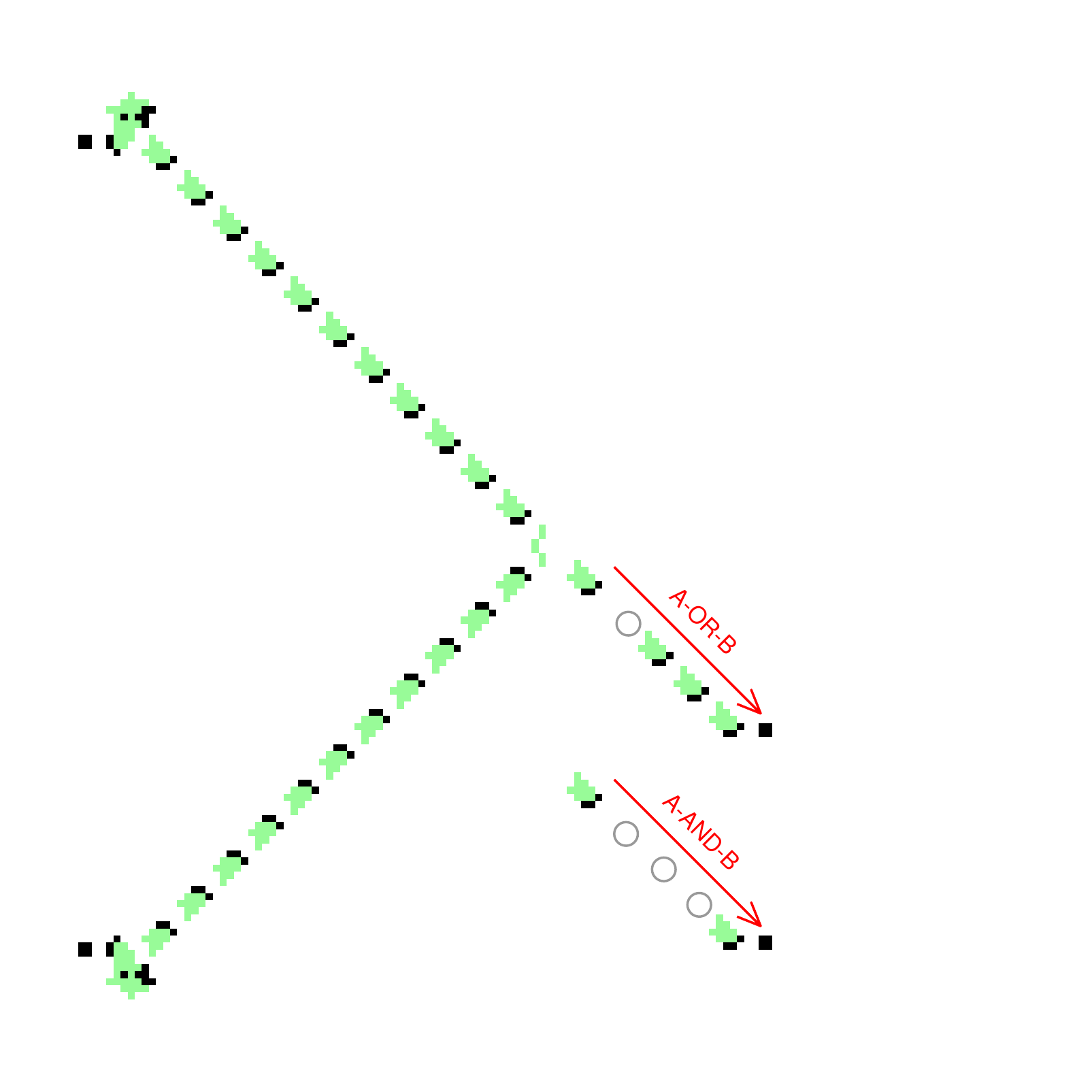}}
\end{minipage}
\end{center}
\vspace{-3ex}
\caption[OR gate]
{\textsf{An example of the OR gate (1 $\vee$ 1 $\rightarrow$ 1, else
    $\rightarrow$ 0) which makes a disjuntion between two stream of data
    represented by two streams of gliders and gaps.
The 5-bit input strings A (11001) and B (10101) both moving SE
interact with two GG1 glider-streams, the lower GG1 shooting NE, and
subsequently with an upper GG1 shooting SE, finally resulting in the A-OR-B
string (11101) moving SE shown after 232 time-steps. The dynamics first makes an
intermediate NOT-A string 00110 (as in figure~\ref{snot1A+}), which
then interacts with string B to simultaneously produce both the AND
string (10001, which appears in the figure) and an intermediate
A-NOR-B string 00010 --- this is inverted by the upper glider-gun 
stream to make NOT(A-NOR-B) which is the same as the A-OR-B string (11101).}}
\label{sor1A+}
\end{figure}

These ingredients exist in Sayab-rule dynamics, where collision
outcomes depend on the precise timing and point of impact.
Interacting GG1 glider-gun streams with glider/gap sequences with the correct
spacing and phases representing a ``string'' of data,
we present examples of the logical gates NOT, AND and OR,
in figures~\ref{snot1A+}, \ref{sand1A+} and \ref{sor1A+}. 
Gaps in a string are indicated by grey circles, and dynamic trails of 10 time-steps
are included. Any input strings can be substituted for those shown. 
Eaters are positioned to eventually stop gliders. 

\section{Concluding remarks}
\label{Concluding remarks} 

The Sayab-rule's glider-gun is the smallest reported to date in 2D CA,
consisting of just four live cells at its minimal phases. From this
glider-gun and other artefacts it is possible to build the logical gates
NOT, AND and OR required for logical universality, which are
constructed by collision dynamics depending on precise timing and
points of impact. Furthermore, the fact that the glider-gun can result
from a collision between two gliders, or between a glider and a simple
oscillator, opens up possibilities for making complex dynamical
structures.

Three basic existential ingredients are proposed for constructing
logical gates, to summarise: a glider-gun, an eater, and
self-destruction when two gliders collide at an angle. Rules with
these ingredients are certainly elusive; in previous
work\cite{Wuensche99,Gomez2015,Gomez2017} we described how they can
nevertheless be found. These methods and the frequency of such rules
in rule-space requires further research. The rules occur as families
of genetically related rules --- this aspect in itself requires
investigation --- for example, variants of the Sayab-rule make up a
family with related behaviour.

Finally, the minimal size of the Sayab-rule's glider-gun is
significant because it should make it easier to interpret its
dynamical machinery, employing De~Bruijn diagrams and other
mathematical and computational tools. Such further research holds the
promise of understanding how glider-guns and related artefacts can
exist, and so reveal the underlying principles of self-organisation in
CA, and by extension in nature itself.

\section{Acknowledgements}
\label{Acknowledgements}

Experiments were done with Discrete Dynamics Lab
\cite{Wuensche2016,Wuensche-DDLab}, Mathematica and Golly.  The Sayab-Rule was
found during a collaboration at June workshops in 2017 at
the DDLab Complex Systems Institute in Ariege, France, and also at the
Universidad Aut\'onoma de Zacatecas, M\'exico, and in London, UK.
J. M. G\'omez Soto acknowledges his residency at
the DDLab Complex Systems Institute, and financial support from the
Research Council of Zacatecas (COZCyT).


\begin{thebibliography}{99}

\begin{small}
  
\bibitem{Banks71}
Banks E,R., ``Information Processing and Transmission in Cellular Automata'', Ph.D. Theses, MIT, Departarment of Mechanical Engineering (1971).

\bibitem{Banks70} Banks, E,R., ``Cellular Automata''. AI Memo No. 198, MIT Artificial Intelligence Lab, 545 Technology SquareÑRoom 821, Cambridge, Massachusetts 02139, 1970.

\bibitem{Berlekamp1982} Berlekamp E,R., J.H.Conway,  R.K.Guy,  
``Winning Ways for Your Mathematical Plays'', 
Vol 2. Chapt 25 ``What is Life?'', 817-850, Academic Press, New York, 1982.

\bibitem{Byl89} Byl, J., ``Self-Reproduction in Small Cellular Automata'',
 Physica D, 34: 295–299, 1989. 

\bibitem{Codd68}
Codd, E., ``Cellular Automata'', Academic Press, (1968).
  
\bibitem{Cook2004} Cook, M.,``Universality in elementary cellular automata''
\textit{Complex Systems}, Vol.15, 1-40, 2004.

\bibitem{Eppstein2010}
Eppstein, D.,``Growth and Decay in Life-Like Cellular Automata'',
in \textit{Game of Life Cellular Automata}, edited by Andrew
Adamatzky, Springer Verlag, 2010.

\bibitem{Francis90}
Pelletier, F.J., and N.M. Martin,
``Post's Functional Completeness Theorem'', \textit{Notre Dame Journal of Formal Logic}, 
Vol.31, No.2, 1990.

\bibitem{Gardner1970} Gardner, M., ``Mathematical
 Games, The fantastic combinations of John Conway's new solitaire
  game "life"''. Scientific American 223. pp. 120--123, 1970.
    
\bibitem{Gomez2015}  G\'omez Soto, J.M., and A.Wuensche,
``The \mbox{X-rule:} universal computation in a
non-isotropic Life-like Cellular Automaton'',
  Journal of Cellular Automata, Vol.10, No.3-4, 261--294, 2015.\\ 
   preprint: \url{http://arxiv.org/abs/1504.01434/}
  
\bibitem{Gomez2017}
G\'omez Soto, J.M., and A.Wuensche, ``X-Rule's Precursor is also Logically Universal'',
\textit{Journal of Cellular}), Vol.12. No.6, 445--473, 2017.\\
preprint: \url{http://arxiv.org/abs/1504.01434/}

\bibitem{Langton84} Langton, C. G., ``Self-reproduction in cellular automata'',
\textit{Physica D}, Vol.10, Issues 1-2, 135--144, 1984.

\bibitem{Langton90} Langton, C. G., ``Computation at the Edge of Chaos'', 
\textit{Physica D}, Vol.42, Issues 1-3, 12--37, 1990.

\bibitem{Post41} Post, E., ``The Two-Valued Iterative Systems of Mathematical Logic'', 
\textit{Annals of Mathematics Series 5}, Princeton University Press, Princeton, NJ, 1941.

\bibitem{Randall2002}
Randall, J-P., ``Turing Universality of the Game of Life'', 
\textit{Collision-Based Computing}, Andrew Adamatzky Ed. Springer Verlag, 2002.

\bibitem{Sapin2004} Sapin,E, O. Bailleux, J.J. Chabrier, and P. Collet.
``A new universal automata discovered by evolutionary algorithms'', 
\textit{Gecco2004.Lecture Notes in Computer Science}, 3102:175--187, 2004.

\bibitem{Sapin2010} Sapin,E, A.Adamatzky, P.Collet, L.Bull, 
``Stochastic automated search methods in cellular automats: the
discovery of tens of thousands of glider guns'',
\textit{Natural Computing} 9:513--543, 2010.

\bibitem{von Neumann}
von Neumann, J., Burks, A. W. (1966), ``Theory of Self-Reproducing Automata'', 
University of Illinois Press, 1966.

\bibitem{Wuensche92} Wuensche,A., and M.Lesser, ``The global Dynamics of
Cellular Automata'', Santa Fe Institute Studies in the Sciences of Complexity,
Addison-Wesley, Reading, MA, 1992.

\bibitem{Wuensche99} Wuensche,A., ``Classifying Cellular Automata
  Automatically; Finding gliders, filtering, and relating space-time patterns,
  attractor basins, and the {\it Z} parameter'', COMPLEXITY, Vol.4/no.3, 47--66, 1999.
  
\bibitem{Wuensche05} Wuensche,A., ``Glider Dynamics in 3-Value Hexagonal
  Cellular Automata: The Beehive Rule'', 
\textit{Int. Journ. of Unconventional Computing}, Vol.1, No.4, 2005, 375-398, 2005.

\bibitem{Wuensche2006} Wuensche,A., A.Adamatzky, ``On spiral
  glider-guns in hexagonal cellular automata: activator-inhibitor
  paradigm'', \textit{International Journal of Modern Physics C}, Vol.17, 
No.7, 1009--1026, 2006.

\bibitem{Wuensche-3D-GG} Wuensche,A., web page: Glider-Guns in 3d Cellular Automata, 2009.\\
\url{http://uncomp.uwe.ac.uk/wuensche/multi_value/3d_glider_guns.html}

\bibitem{Wuensche2016}
Wuensche,A., ``Exploring Discrete Dynamics - Second Edition'', Luniver Press, 2016.

\bibitem{Wuensche-DDLab}  Wuensche,A., Discrete Dynamics Lab (DDLab), 1993-2017.\\
  \url{http://www.ddlab.org/}
\end{small}

\end{thebibliography}
\end{document}